\documentclass[%
 reprint,
 amsmath,amssymb,
 longbibliography,
 aps,
 pre,
]{revtex4-1}

\usepackage[normalem]{ulem}
\usepackage{graphicx}
\usepackage{bm}
\usepackage{xcolor}
\usepackage{hyperref}
\usepackage{amsmath}
\usepackage[utf8]{inputenc}
\usepackage{ulem}
\usepackage{mathtools}
\usepackage{xr}

\newcommand{\mean}[1]{\mathbb{E}\left[#1\right]}
\newcommand{\var}[1]{\mathrm{Var}\left(#1\right)}
\newcommand{\varasy}[1]{\mathrm{Var}^{\mathrm{Asy}}\!\!\left(#1\right)}
\newcommand{\varnum}[1]{\mathrm{Var}^{\mathrm{Num}}\left(#1\right)}
\newcommand{\meanasy}[1]{\mathbb{E}^{\mathrm{Asy}}\left[#1\right]}
\newcommand{\meannum}[1]{\mathbb{E}^{\mathrm{Num}}\left[#1\right]}

\def\env{\mathrm{Env}_L^N}
\def\sam{\mathrm{Sam}_L^N}
\def\min{\mathrm{Min}_L^N}
\def\minSSRW{\mathrm{\widetilde{Min}}_L^N}
\def\ISSRW{\tilde{I}}
\def\tauL{\tau_L}
\def\pdf{p_{L}^{\bold{B}}}
\def\rwre{RWRE}

\makeatletter
\newcommand{\hathat}[1]{%
\begingroup%
  \let\macc@kerna\z@%
  \let\macc@kernb\z@%
  \let\macc@nucleus\@empty%
  \hat{\raisebox{.2ex}{\vphantom{\ensuremath{#1}}}\smash{\hat{#1}}}%
\endgroup%
}
\makeatother

\begin{document}

\title{First Passage Time for Many Particle Diffusion in Space-Time Random Environments
}
\author{Jacob B. Hass$^*$, Ivan Corwin$^\dagger$, Eric I. Corwin$^*$}
\affiliation{$^*$Department of Physics and Materials Science Institute, University of Oregon, Eugene, Oregon 97403, USA. \\ $^\dagger$Department of Mathematics, Columbia University, New York, New York 10027, USA.}
\date{\today}

\begin{abstract}
The first passage time for a single diffusing particle has been studied extensively, but the first passage time of a system of many diffusing particles, as is often the case in physical systems, has received little attention until recently. We consider two models for many particle diffusion -- one treats each particle as independent simple random walkers while the other treats them as coupled to a common space-time random forcing field that biases particles nearby in space and time in similar ways. The first passage time of a single diffusing particle under both of these models show the same statistics and scaling behavior. However, for many particle diffusions, the first passage time among all particles (the `extreme first passage time') is very different between the two models, effected in the latter case by the randomness of the common forcing field. We develop an asymptotic (in the number of particles and location where first passage is being probed) theoretical framework to separate out the impact of the random environment with that of sampling trajectories within it. We identify a new power-law describing the impact to the extreme first passage time variance of the environment. Through numerical simulations we verify that the predictions from this asymptotic theory hold even for systems with widely varying numbers of particles, all the way down to 100 particles. This shows that measurements of the extreme first passage time for many-particle diffusions provide an indirect measurement of the underlying environment in which the diffusion is occurring.
\end{abstract}

\maketitle

\section{Introduction}
The \emph{extreme} (or fastest) first passage time beyond a barrier among many particles diffusing in a common environment determines the function of a variety of system such as oocyte fertilization \cite{meersonMortalityRedundancyDiversity2015, schussRedundancyPrincipleRole2019, yangSearchSmallEgg2016}, neuronal activation \cite{basnayakeFastCalciumTransients2019} and information flow in networks \cite{germanoTrafficParticlesComplex2006, wangFirstPassageTime2008}. When particles are modeled as independent simple symmetric random walks (SSRW) or Brownian motions \cite{einsteinZurTheorieBrownschen1906, einsteinUberMolekularkinetischenTheorie1905, einsteinTheoretischeBemerkungenUber1907}, there has been extensive study of the first passage behavior for a single particle \cite{rednerGuideFirstPassageProcesses2001a}, or recently for many particles \cite{meersonMortalityRedundancyDiversity2015, linnExtremeHittingProbabilities2022,schussRedundancyPrincipleRole2019,  rednerRedundancyExtremeStatistics2019, lawleyDistributionExtremeFirst2020, lindenbergLatticeRandomWalks1980, madridCompetitionSlowFast2020,grebenkovSingleparticleStochasticKinetics2020,lawleyUniversalFormulaExtreme2020}.

We probe the behavior of extreme first passage time for a model of many particle diffusion where particles are modeled as random walks in a common environment of space-time inhomogeneous biases which are, themselves, modeled by a random field with short-range correlations in space and time. We focus on this (time-dependent) random walk in random environment (\rwre{}) model in one spatial dimension as would be relevant to diffusion in long and thin capillaries. Self-averaging of the environment implies that the statistical behavior of a single (or typical) particle in the \rwre{} model remains unchanged compared to the SSRW model \cite{rassoul-aghaAlmostSureInvariance2005, deuschelQuenchedInvariancePrinciple2018, rassoul-aghaQuenchedFreeEnergy2013}. However, the extreme behavior of particles in the \rwre{} model has recently been shown to be quite different to that of the SSRW, with statistics and power-laws related to the Kardar-Parisi-Zhang (KPZ) universality class and equation \cite{barraquandRandomwalkBetadistributedRandom2017, barraquandModerateDeviationsDiffusion2020, hassAnomalousFluctuationsExtremes2023} (see also \cite{thieryExactSolutionRandom2016, corwinKardarParisiZhang2017a, ledoussalDiffusionTimedependentRandom2017a,barraquandLargeDeviationsSticky2020, barraquandRandomWalkNonnegative2023, brockingtonBetheAnsatzSticky2023, brockingtonEdgeCloudBrownian2022, oviedoSecondOrderCubic2022, dasKPZEquationLimit2023, hartmannProbingLargeDeviations2023, krajenbrinkCrossoverMacroscopicFluctuation2023}). Those works focus on the statistical behavior of the locations of the particles that move the furthest  from a common starting position as a function of time and the number of particles, $N$. The location of the furthest particle is a complementary measurement to the extreme first passage time which measures time as a function of the location of a boundary.

Here we leverage the theoretical and asymptotic (in $N\to \infty$) results on extreme particle locations in the \rwre{} model \cite{barraquandRandomwalkBetadistributedRandom2017, barraquandModerateDeviationsDiffusion2020} to make precise finite $N$ predictions about extreme first passage time statistics as a function of the location, $L$, of first passage and the number of particles, $N$. We show that the effect of the randomness of the environment and of sampling $N$ random walks in that environment are approximately independent. This means that by probing the extreme first passage time we are able to gain access to certain measurements of the hidden environment in which the diffusion occurs. Using numerical simulations, we show that these predictions remain valid down to quite small systems of $N=100$.

\section{Background}
Classical modeling of diffusion \cite{einsteinZurTheorieBrownschen1906, einsteinUberMolekularkinetischenTheorie1905, einsteinTheoretischeBemerkungenUber1907} has served as a basis for the development of more complex models such as L\'evy flights in biological systems \cite{wangWhenBrownianDiffusion2012}, anomalous diffusion where the mean squared distance is not linear \cite{metzlerBrownianMotionFirstpassage2019, bouchaudAnomalousDiffusionDisordered1990}, and active materials where particles have internal stores of energy \cite{ramaswamyMechanicsStatisticsActive2010, kanazawaLoopyLevyFlights2020}. The aforementioned models modify the existing framework of classical diffusion to better model specific phenomenon, whereas we study a model that aims to better capture the behavior of classical diffusion in generality.

We consider a particular type of random walk in random environment (\rwre{}) models. \rwre{} models have been studied extensively and come in various forms, including those with short range correlated forces \cite{richardsonAtmosphericDiffusionShown1926, hentschelRelativeDiffusionTurbulent1984, bouchaudDiffusionLocalizationWaves1990,chertkovAnomalousScalingExponents1996, bernardAnomalousScalingNPoint1996}, long range correlated forces \cite{bouchaudAnomalousDiffusionDisordered1990, kestenLimitLawRandom1975, sinaiLimitingBehaviorOneDimensional1983, bouchaudClassicalDiffusionParticle1990, burlatskyTransientRelaxationCharged1998, kestenLimitLawRandom1975} and only spatially varying random forces \cite{chernovReplicationMulticomponentChain1967, temkinOnedimensionalRandomWalks1972, kestenLimitLawRandom1975, kestenLimitLawRandom1975, kestenLimitLawRandom1975, hughesRandomWalksRandom1995, oferRandomWalksRandom2004}. We consider a short range, spatially and temporally varying random field.

Extreme first passage time statistics for classical models of many particle diffusion (e.g. $N$ independent SSRWs) have been studied extensively \cite{meersonMortalityRedundancyDiversity2015, linnExtremeHittingProbabilities2022,schussRedundancyPrincipleRole2019,  rednerRedundancyExtremeStatistics2019, lawleyDistributionExtremeFirst2020, lindenbergLatticeRandomWalks1980, madridCompetitionSlowFast2020,grebenkovSingleparticleStochasticKinetics2020,lawleyUniversalFormulaExtreme2020}. For \rwre{} models with a temporally constant environment \cite{ledoussalFirstpassageTimeRandom1989, lawleyUniversalFormulaExtreme2020, noskowiczAverageTypicalMean1988} studied the aspects of first passage times for one and many particle diffusions. In our setting of a temporally varying environment, \cite{ledoussalDiffusionTimedependentRandom2017a} (see the supplementary materials in \cite{doussalDiffusionTimedependentRandom2017a}) initiated the study of extreme first passage time upon which we will expand.

\section{Model for Diffusion}\label{sec:model}
We consider independent random walks subject to a common environment which determines the bias at each site. We model the environment as independent and identically distributed (i.i.d.) transition biases $\bold{B}=\{B(x,t): x\in \mathbb{Z}, t \in \mathbb{Z}_{\geq0} \}$ drawn from a distribution supported on $[0, 1]$. In this work we focus on the \emph{Random Walk in Random Environment (\rwre{})} model where $B(x,t)$ are drawn from the uniform distribution on $[0, 1]$. If instead all $B(x, t)= 1/2$ our model reduces to the \emph{Simple Symmetric Random Walk (SSRW)} model for diffusion. We write $\mathbb{P}(\bullet)$ for the probability of an event $\bullet$, and $\mathbb{E}[\bullet]$ and $\var{\bullet}$ for the expectation and variance of a random variable $\bullet$ averaged over the random environment $\bold{B}$. For a given environment $\bold{B}$, we use that notation $\mathbb{P}^{\bold{B}}(\bullet)$, $\mathbb{E}^{\bold{B}}[\bullet]$ and $\mathrm{Var}^\bold{B}(\bullet)$ to represent the probability of an event $\bullet$ (in the first case) or random variable $\bullet$ (in the second and third cases) given the environment $\bold{B}$. Recall the laws of total probability, expectation and variance
\begin{equation}\label{eq:total}
\begin{aligned}
\mathbb{P}(\bullet) &= \mathbb{P}(\mathbb{P}^{\bold{B}}(\bullet)),\qquad \mathbb{E}[\bullet] = \mathbb{E}[\mathbb{E}^{\bold{B}}[\bullet]], \\
\var{\bullet} &= \var{\mathbb{E}^{\bold{B}}[\bullet]} + \mathbb{E}[\mathrm{Var}^{\bold{B}}(\bullet)].
\end{aligned}
\end{equation}

Let us describe now the motion of a single particle given the environment $\bold{B}$. We denote the position of a single particle at time $t \in \mathbb{Z}_{\geq 0}$ as $X(t) \in \mathbb{Z}$. It evolves as the following Markov chain. We begin at the origin such that $X(0)=0$. Subsequently for all $t \in \mathbb{Z}_{\geq 0}$ if the  particle is at position $X(t)=x$ it flips a weighted coin which has probability of heads $B(x,t)$ and tails $1-B(x,t)$. If heads, the particle changes position such that $X(t+1) = X(t) + 1$ and if tails then $X(t+1) = X(t) - 1$. We use $p^\bold{B}(x, t) := \mathbb{P}^\bold{B}(X(t)=x)$ to denote the probability mass function for $X(t)$ which uniquely solves the Kolmogorov backwards equation
\begin{equation*}
\begin{aligned}
p^\bold{B}(x, t) = &p^{\bold{B}}(x-1, t-1) B(x-1, t-1)\\ &+ p^{\bold{B}}(x+1, t-1)\big(1-B(x+1, t-1)\big)
\end{aligned}
\end{equation*}
for $x \in \mathbb{Z}$ and $t \in \mathbb{Z}_{\geq 0}$
with initial condition $p^\bold{B}(x, 0) = \mathbf{1}_{x=0}$ (with the notation $\mathbf{1}_{E}$ equals $1$ if the event $E$ occurs and 0 otherwise). For  $L \in \mathbb{N}$ we define the (random) first passage time, $\tauL$, for $X(t)$ as
$$
\tau_L = \mathrm{min}\big(t: X(t) \notin (-L, L)\big),
$$
i.e., the time when the particle first exits $(-L, L)$.

\begin{figure}
\includegraphics[width=\columnwidth]{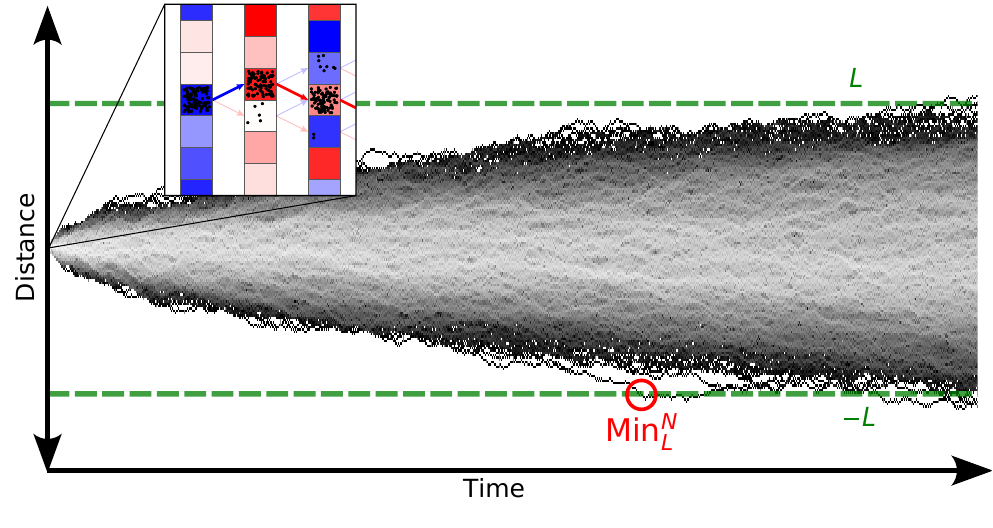}
\caption{A system of $N=10^5$  particles evolving according to our RWRE model. The green dashed lines denote a barrier at $-L$ and $L$. The extreme first passage time, $\min$, is identified by the red circle and is the first time when one of these $N$ walkers crosses this barrier. The inset illustrates the dynamics of particles in the first three time steps (not all $10^5$ particles are shown to avoid cluttering). The color of each box in the inset corresponds to the bias of the location (blue means upward bias and red means downward bias).}
\label{fig:Occupation}
\end{figure}

We consider $N$ particles $X^1(t),\ldots, X^N(t)$ evolving independently in the same environment $\bold{B}$. This means that if multiple particles are at the same location and the same time, they use the same biased coins to determine their next move, though the flips of those coins are done independently of each other. For $i\in \{1,\ldots, N\}$ let $\tauL^i$ denote the first passage time for particle $X^i(t)$. Given the environment, the laws of $\tauL^1,\ldots, \tauL^N$ are independent and identically distributed.
We will study the first time {\it any} of the $N$ particles leave $(-L, L)$ which we call the {\it extreme first passage time} and denote by $\min = \mathrm{min}(\tauL^1,...,\tauL^N)$. A visualization of our system and the extreme first passage time can be seen in Fig. \ref{fig:Occupation}.

The probability distribution of $\tauL$ and consequently  that of $\min$ can be computed by studying $X_L(t) := X(\mathrm{min}(t, \tauL))$, the random walk $X(t)$ stopped (or absorbed) when it first exits $(-L, L)$. The corresponding probability mass function $\pdf(x,t) := \mathbb{P}^{\bold{B}}(X_L(t) = x)$  uniquely solves the Kolmogorov backwards equation
\begin{equation}\label{eq:MasterEq}
\begin{aligned}
\pdf(x, t) = &\pdf(x-1, t-1) B(x-1, t-1)\\ &+ \pdf(x+1, t-1)\big(1-B(x+1, t-1)\big)
\end{aligned}
\end{equation}
for $x \in [-L+2,L-2]\cap \mathbb{Z}$ and $t \in \mathbb{Z}_{\geq 0}$ subject absorbing boundary conditions whereby
\begin{equation}
\begin{aligned}\label{eq:bcs}
\pdf(L, t) &= \pdf(L, t-1) \\&+ B(L-1, t-1) \pdf(L - 1, t-1), \\
\pdf(L-1, t) &=\pdf(L-2, t-1) B(L-2, t-1), \\
\pdf(-L+1, t)&= \pdf(-L+2, t-1)\big(1-B(-L+2, t-1)\big),\\
\pdf(-L, t) &= \pdf(-L, t-1) \\&+ \pdf(-L+1, t-1)\big(1-B(-L+1, t-1)\big),
\end{aligned}
\end{equation}
and initial condition
$\pdf(x, 0) = \mathbf{1}_{x=0}$.
The probability of $\tauL$ occurring before time $t$ is the same as the probability of being absorbed before time $t$ which is given by
\begin{equation}\label{eq:FPTSpatial}
\mathbb{P}^{\bold{B}}(\tauL \leq t) = \pdf(L,t) + \pdf(-L, t).
\end{equation}
From this, the probability distribution of the extreme first passage time $\min$ is then given by
\begin{equation}\label{eq:NFPT}
\mathbb{P}^{\bold{B}}(\min\leq t) = 1 - \big(1-\mathbb{P}^{\bold{B}}(\tauL \leq t)\big)^N
\end{equation}
since each $\tauL^i$ is independent and identically distributed with probability distribution function $\mathbb{P}^{\bold{B}}(\tauL \leq t)$.

There are two sources of randomness at play in $\min$. The first is due to the randomness of the underlying environment $\bold{B}$ and the second due to sampling random walks in said environment. We seek to study the impact of each and their interplay. We define a natural proxy, $\env$, for the randomness due to the environment by
\begin{equation}\label{eq:EnvDef}
\env \coloneqq \mathrm{min}\left(t: \mathbb{P}^{\bold{B}}(\tauL \leq t) \geq \frac{1}{N}\right).
\end{equation}
Notice that $\env$ is deterministic given $\bold{B}$ and, in light of Eq. \ref{eq:NFPT}, that it approximately satisfies
$\mathbb{P}^{\bold{B}}(\min\leq \env) \approx 1-(1-1/N)^N\approx e^{-1}$. The reason the first approximation is not an equality is because generally $\mathbb{P}^{\bold{B}}(\tauL \leq t) \geq \frac{1}{N}$ in Eq. \ref{eq:EnvDef} will not strictly be equal to $1/N$. The difference
\begin{equation}\label{eq:SamDef}
\sam \coloneqq\min - \env,
\end{equation}
which is still random given $\bold{B}$, contains the randomness from sampling $\tauL^1,\ldots, \tauL^N$ given the environment $\bold{B}$.

While we have chosen to study the extreme first passage time of one or many particles leaving the region $(-L, L)$, we could just as well have studied the analogous time for leaving the region $(-\infty, L)$. The former case (studied mostly here) is the two-sided case while the latter is the one-sided case. The two-sided case benefits from the fact that $\mathbb{E}[\min]<\infty$ for all $N\geq 1$ while in the one-sided case, this mean is infinite for small enough $N$. Owing to this heavy-tailed nature of the one-sided case, it is numerically less efficient to study than in the two-sided case. On the other hand, our theoretical framework and asymptotic predictions provided below can be easily extended to the one-sided case. We will mostly focus on the two-sided case but also record some results for the one-sided case. Thus, in anticipation of that, let us define the first passage time $\tilde{\tau}_L = \mathrm{min}(t: X(t) < L)$ and the stopped random walk $\tilde{X}_L(t) := X(\mathrm{min}(t, \tilde{\tau}_L))$, whose probability mass function $\tilde{p}_L^{\bold{B}}(x,t) := \mathbb{P}^{\bold{B}}(\tilde{X}_L(t) = x)$ obeys the same equation and initial condition as the double sided case, but without the absorbing boundary  at $-L$. Thus $\tilde{p}_L^{\bold{B}}(x,t)$ satisfies Eq. \ref{eq:MasterEq} for $x \in (-\infty,L-1)\cap \mathbb{Z}$ subject to the first two equalities in Eq. \ref{eq:bcs}.
\section{Numerical Methods}
In what follows we will study the mean and variance of $\min,\env$ and $\sam$. We will numerically measure the values of these quantities via methods described here, and to distinguish these numerical measurements from the true value, we will introduce a superscript so that $\meannum{\bullet}$ and $\varnum{\bullet}$ represent the numerically measured mean and variances of $\bullet$.

Given an environment $\bold{B}$, we can exactly sample  the motion of $N$ particles in it by following the agent-based definition whereby each particle evolves as a random walk with biases given by $\bold{B}$. When $N$ is large it is not feasible to use the agent-based approach.  Instead, as in \cite{hassAnomalousFluctuationsExtremes2023}, we obtain massive increase in efficiency by noting that if there are $N(x,t)$ particles at site $x$ and time $t$, they will split into right and left moving populations according to a Binomial distribution. Specifically, the number of particles that move from site $x$ at time $t$ to site $x+1$ at time $t+1$ is drawn from a Binomial distribution with $N(x,t)$ trials and success probability $B(x,t)$. The remaining particles move to site $x-1$ at time $t+1$. This occupation variable based approach provides a way to exactly sample the number of particles per site over time which is sufficient to study the first passage times in question here.

In this work we present results on systems of $N=10^2$ to $N=10^{28}$ particles and measure the time of first passage for distances up to $L=750\cdot\ln(N)$. We measure the extreme first passage time past multiple distances in the same simulation by recording the time at which a particle first leaves each given boundary. We numerically measure $\mean{\min}$ and $\var{\min}$ by choosing, according to the uniform distribution on each $B(x,t)$, an environment, $\bold{B}$, and then sampling $\min$ as described above. Repeating this for many independently sampled environments allows us to build up the distribution of $\min$ and thus estimate $\mean{\min}$ and $\var{\min}$. For different values of $N$ we repeat this procedure. However, for a given value of $N$, we make use of the same environment to study the statistics of $\min$ for multiple values of $L$. Though this introduces some correlation in these numerically computed statistics at different distances, these should become negligible for a large sample size (i.e., the number of different instances of $\bold{B}$ used).

To study $\env$, we compute $\pdf(x,t)$ for a given environment, $\bold{B}$, and distance $L$ by numerically solving Eq. \ref{eq:MasterEq} with the boundary conditions Eq. \ref{eq:bcs} and initial condition $\pdf(x, 0) = \mathbf{1}_{x=0}$. From $\pdf(x,t)$ we calculate $\mathbb{P}^{\bold{B}}(\tauL \leq t)$ using Eq. \ref{eq:FPTSpatial}. We then measure $\env$ for a given environment using its definition in Eq. \ref{eq:EnvDef}. By computing $\mathbb{P}^{\bold{B}}(\tauL \leq t)$ for many independent samples of the environment $\bold{B}$, we estimate  $\mean{\env}$ and $\var{\env}$.

We measure $\sam$ using the following process. For a given environment $\bold{B}$, we calculate the probability distribution of $\sam$ using Eq. \ref{eq:NFPT} and \ref{eq:SamDef}, i.e.,
\begin{equation}
\mathbb{P}^\bold{B}(\sam \leq t) = 1 - \big(1 - \mathbb{P}^\bold{B}(\tauL \leq t + \env)\big)^N.
\end{equation}
Using $\mathbb{P}^\bold{B}(\sam \leq t)$, we calculate $\mathbb{E}^\bold{B}[\sam]$ and $\mathrm{Var}^\bold{B}[\sam]$ numerically. We are using the notation $\mathbb{E}^{\bold{B}}[\bullet]$ and $\mathrm{Var}^{\bold{B}}[\bullet]$ introduced in Sec. \ref{sec:model}. We can then numerically measure  $\mathbb{E}[\sam]$ and $\mathrm{Var}[\sam]$ by sampling many independent instances of $\bold{B}$ and using the laws of total expectation and variance in Eq. \ref{eq:total}.

We probe four values of $N$, $N=10^2, 10^5, 10^{12}$ and $10^{28}$. To measure statistics involving $\min$ we use $20,000, 10,000, 10,000$ and  $2,500$ systems for the respective values of $N$, while to measure statistics involving $\env$ and $\sam$ we use $2,000, 2,000, 2,000$ and $1,000$ systems for the respective values of $N$. Notice that we used fewer systems for $\env$ and $\sam$ than for $\min$. This is because $\min$ can be sampling via simulating the motion of $N$ particles as described above while $\env$ and $\sam$ require computing the probability distribution by solving the master equation. The former is computationally much less expensive than the latter, hence our reduction in number of iterations. For measuring statistics involving $\min$ for the SSRW model we use $5,000$ systems (in other words, we repeatedly sampling $N$ SSRWs and observe $\min$ for a total of $5,000$ instances).

To compute the data presented in this paper, we used 500 CPUs on a high performance computing cluster for approximately two weeks. Our code is available at \url{https://github.com/CorwinLab/RWRE-Simulations}.

\section{Derivation of Asymptotic Predictions}

We derive asymptotic predictions for the extreme first passage time for the \rwre{} model in the limit where both $L$ and $N$  tending towards infinity with certain relationships. We study the mean and variances of $\min,\env$ and $\sam$ and develop asymptotic formulas  that we subsequently compare to our numerical simulations in Sec. \ref{sec:compare}. We use the notation $\meanasy{\bullet}$ and $\varasy{\bullet}$ to   distinguish our formulas from the true values of $\mean{\bullet}$ and $\var{\bullet}$ in question.

We find that there are three different scaling regimes with smooth transitions between them, consistent with previous results in \cite{hassAnomalousFluctuationsExtremes2023} (which probes for different $N$ the location of the maximum as a function of time, instead of the first passage times as a function of barrier location). The short distance regime is when $L/\ln(N)\to \hat{L}<1$, in which case we will easily see that with very high probability at least one particle will move ballistically (hence resulting in trivial behaviors for the mean and variances in question). The medium distance regime is when $L/\ln(N)\to \hat{L}\in (1,\infty)$ in which case we leverage results from \cite{barraquandRandomwalkBetadistributedRandom2017} to derive our asymptotic formulas related to the Gaussian Unitary Ensemble (GUE) Tracy-Widom (TW) distribution \cite{tracyLevelspacingDistributionsAiry1994}. The large distance regime is when $L/(\ln(N))^{3/2}\to \hathat{L}\in (0,\infty)$ in which case we leverage results from \cite{barraquandModerateDeviationsDiffusion2020} to derive our asymptotic formulas related to the statistics of the solution to the KPZ equation with narrow wedge initial data \cite{kardarDynamicScalingGrowing1986,sasamotoOneDimensionalKardarParisiZhangEquation2010,calabreseFreeenergyDistributionDirected2010,dotsenkoBetheAnsatzDerivation2010,amirProbabilityDistributionFree2011}.
In each of these regimes we derive asymptotic approximations for the mean and variance of $\sam$ and $\env$ and argue they are independent when $L$ and $N$ tend towards infinity. Using this independence and these asymptotics for $\sam$ and $\env$, we then likewise provide formulas for the asymptotic mean and variance of the extreme first passage time, $\min$, as recorded in Eqs. \ref{eq:varasymin} and \ref{eq:meanasymin}.

The study of extreme first passage time for the \rwre{} model was initiated in work of \cite{ledoussalDiffusionTimedependentRandom2017a} (in particular, see the supplemental material in \cite{doussalDiffusionTimedependentRandom2017a}) which focused on the fluctuations of $\env$ in the medium distance regime. Our work in this section builds upon that analysis, offering some additional justification (based in part on KPZ scaling theory) and refinement as well as expanding to the large distance regime. We also develop the theory describing the sampling fluctuations $\sam$. In particular, we describe how the environmental and sampling fluctuations should be independent. Moreover, we propose a finite $N$ formula based on stitching together these two asymptotic regimes. In the subsequent Section \ref{sec:compare} we verify our theory for a wide range of $N$ through numerical simulations.

\subsection{Short Distance Regime}
We assume that $L,N\to \infty$ with $L/\ln(N)\to \hat{L}<1$. In this case, it is very likely that at least one of the $N$ particles will arrive at position $\pm L$ at time $t=L$ in which case we can conclude that
$\meanasy{\min}= L$  and  $\varasy{\min}= 0$.
Similarly, we see that $\meanasy{\env}= L$ and $\varasy{\env}=\meanasy{\sam}=\varasy{\sam}=0$.
To see this observe that the probability that a given particle will arrives at position $\pm L$ at time $t=L$ is given by
$$
p^{\bold{B}}(L, L) + p^{\bold{B}}(-L, L) = \prod_{x=1}^L B(x,x)+ \prod_{x=1}^{-L} \big(1- B(x,-x)\big)
$$
where the $B(x,t)$ are i.i.d. uniform random variables. We use the law of large numbers to see that
$$
\ln\big(p^{\bold{B}}(L, L)\big) = \sum_{x=1}^{L} \ln\big(B(x,x)\big) \approx -L
$$
where we have used the fact that $-\ln(B(x,x))$ is an  exponential mean 1 random variable. The same holds for $ p^{\bold{B}}(-L, L)$ thus we see that
$$
p^{\bold{B}}(L, L) + p^{\bold{B}}(-L, L)  \approx 2 e^{-L}\approx 2 e^{-\hat{L} \ln(N)}\gg 1/N
$$
where we have used that $L/\ln(N)\to \hat{L}<1$.
When $p^{\bold{B}}(L, L) + p^{\bold{B}}(-L, L) \gg 1/N$ (as above) it is likely that at least one of $N$ independent particles is at $\pm L$. This implies the above claimed results when $\hat{L}<1$.

\medskip
\noindent{\it One-sided case:} In the one-sided barrier case the exact same analysis above goes through except only the $ p^{\bold{B}}(L, L)$ term should be considered above.

\subsection{Medium Distance Regime $\env$ Behavior}
The behavior in the medium and large distance regimes is considerably more complex. We first study the random variable $\env$ in these regimes. Then, based on asymptotics derived for its mean and variance we determine asymptotic predictions for $\sam$ and $\min$.

In the medium and large distance regime our analysis relies on results  \cite{barraquandRandomwalkBetadistributedRandom2017, barraquandModerateDeviationsDiffusion2020} derived using tools from quantum integrable systems. We will review these results below. In essence they provide precise asymptotic information about the distribution of  $\mathbb{P}^{\bold{B}}(X(t)\geq L)$ in various limits as $t$ and $L$ grow. This information is not equivalent to the knowledge of the first passage time distribution $\mathbb{P}^{\bold{B}}(\tauL\leq t)$ that we seek to understand here. While the event $\{ X(t)\geq L\}$ implies $\{\tauL\leq t\}$, the opposite is not true. It is possible that a particle will pass the barrier at a time prior to $t$ and then backtrack behind it at time $t$. However, as $L$ approaches $t$ the events become more and more equivalent since it is harder for a particle to exit $(-L,L)$ and then backtrack when $L$ is large. For instance, when $L=t$ (the maximal possible value) the two events become equivalent. Based on this reasoning we make the following approximation that improves as $L$ grows relative to $t$:
$$\pdf(L,t) \approx \mathbb{P}^{\bold{B}}(X(t) \geq L), \quad \pdf(-L, t)\approx \mathbb{P}^{\bold{B}}(X(t) \leq -L).$$
Recall that $ \pdf(L,t)$ and $\pdf(-L,t)$ are the probabilities of absorption at $L$ and $-L$ up to time $t$ and their sum, as in
Eq. \ref{eq:FPTSpatial}, yields $\mathbb{P}^{\bold{B}}(\tauL \leq t)$. Thus, we arrive at the starting approximation for our analysis:
\begin{equation}\label{eq:UnboundedApprox}
\mathbb{P}^{\bold{B}}(\tauL \leq t) \approx \mathbb{P}^{\bold{B}}(X(t) \geq L) + \mathbb{P}^{\bold{B}}(X(t) \leq -L).
\end{equation}

In the medium distance regime we assume that $L,N\to \infty$ with $L/\ln(N)\to \hat{L}\in (1,\infty)$.
The key input from \cite{barraquandRandomwalkBetadistributedRandom2017} is as follows. Define the random variable $\chi_{x, t}$ by
\begin{equation}\label{eq:BC}
 \ln(\mathbb{P}^{\bold{B}}(X(t) \geq x)) = -t I\left(\frac{x}{t}\right) + t^{1/3} \sigma\left(\frac{x}{t}\right) \chi_{x, t}
\end{equation}
where $x\in [0,t]\cap \mathbb{Z}$ and for $v\in [0,1)$
$$I(v) = 1 - \sqrt{1-v^2},\quad \textrm{and}\quad \sigma(v)^{3} = \frac{2I(v)^2}{1-I(v)}.$$
Then \cite{barraquandRandomwalkBetadistributedRandom2017} shows that for $v\in (0,1)$, $\chi_{vt,t}$ converges as $t\rightarrow\infty$ in distribution to a GUE TW random variable. By symmetry the result of \cite{barraquandRandomwalkBetadistributedRandom2017} also holds if in Eq. \ref{eq:BC},
$\ln(\mathbb{P}^{\bold{B}}(X(t) \geq x))$ is replaced by $\ln(\mathbb{P}^{\bold{B}}(X(t) \leq -x))$ and $\chi_{x,t}$ is replaced by $\chi_{-x, t}$.  It should be noted that while \cite{barraquandRandomwalkBetadistributedRandom2017} does not address the joint distribution of $\chi_{x, t}$ for different values of $x$ and $t$, it is possible to make predictions based on grounds of KPZ universality \cite{corwinKardarParisiZhang2012, quastelOneDimensionalKPZEquation2015}. In particular, for any $v\in (-1,0)\cup (0,1)$ as $T \to \infty$, the space-time random process $(x,t)\mapsto \chi_{vtT+xT^{2/3},tT}$ should converge to the KPZ fixed point \cite{matetskiKPZFixedPoint2021}, and the limiting processes for distinct $v$ should be independent. Moreover, the local regularity of the KPZ fixed point is known which should translate into estimates on the regularity of $\chi_{x,t}$. Some of this understanding will justify the approximations that follow involving how $\chi_{x,t}$ changes when $t$ varies.

Combining Eq. \ref{eq:UnboundedApprox} and Eq. \ref{eq:BC} yields
\begin{equation}
\begin{aligned}\label{eq:MedDistanceProb}
&\ln\big(\mathbb{P}^{\bold{B}}(\tauL \leq t)\big) \approx \\
&- t I\left(\tfrac{L}{t}\right) + \ln\Big(e^{t^{1/3}\sigma(\frac{L}{t})\chi_{L, t}} + e^{t^{1/3}\sigma(\frac{L}{t})\chi_{-L, t}}\Big)
\end{aligned}
\end{equation}
We seek to study the random variable $\env$ which is defined by Eq. \ref{eq:EnvDef} and essentially is the $t$ such that $\mathbb{P}_\bold{B}(\tauL \leq t) = 1/N$ holds. This yields the following implicit equation for $\env$
\begin{equation}
\begin{aligned}\label{eq:medLogN}
&-\ln(N) \approx - \env \,\cdot \,I\left(\tfrac{L}{\env}\right) + \\
& \ln\Big(e^{\left(\env\right)^{1/3}\sigma(\frac{L}{\env})\chi_{L, \env}} + e^{\left(\env\right)^{1/3}\sigma(\frac{L}{\env})\chi_{-L, \env}}\Big).
\end{aligned}
\end{equation}
We will solve this equation perturbatively. The first order solution neglects the second line in Eq. \ref{eq:medLogN} and yields \begin{equation}\label{eq:TWFirstOrder}
\env \approx T_0 := \frac{(\ln(N))^2 + L^2}{2 \ln(N)}
\end{equation}
i.e., $T_0$ solves $\ln(N) = T_0 I(\tfrac{L}{T_0})$.
We now assume that $\env \approx T_0 + \delta$ with $\delta \ll T_0$ containing the randomness of $\env$. Substituting this into Eq. \ref{eq:medLogN} we can find $\delta$ approximately by solving
\begin{equation}
\begin{aligned}\label{eq:compars}
-\ln(N)  \approx& -(T_0 + \delta)\,\cdot\, I\left(\tfrac{L}{T_0 + \delta}\right) + \\ &\ln\Big(e^{T_0^{1/3}\sigma\left(\frac{L}{T_0}\right)\chi_{L, T_0}} + e^{T_0^{1/3}\sigma\left(\frac{L}{T_0}\right)\chi_{-L, T_0}}\Big) \\
\approx& -(T_0 + \delta) \,\cdot \,\left( I\left(\tfrac{L}{T_0}\right) + \delta \partial_t\left. I\left(\tfrac{L}{t}\right)\right|_{t=T_0} \right)+ \\
&\ln\Big(e^{T_0^{1/3}\sigma\left(\frac{L}{T_0}\right)\chi_{L, T_0}} + e^{T_0^{1/3}\sigma\left(\frac{L}{T_0}\right)\chi_{-L, T_0}}\Big).
\end{aligned}
\end{equation}
In the first comparison we neglected the fact that under the perturbation to $T_0$ we should write $\chi_{L, T_0+\delta}$ and $\chi_{-L, T_0+\delta}$. This, however, is justified by the fact that while $T_0$ is of order $\ln(N)$, $\delta$ (as we will see below) is of order $(\ln(N))^{1/3}$. By the KPZ scaling theory mentioned earlier, this change in the time variable of the $\chi$ process should have a small impact which we neglect in our approximation. In the second comparison we utilized the Taylor expansion $I\left(\frac{L}{T_0 + \delta}\right) \approx I\left(\frac{L}{T_0}\right) + \delta \partial_t\left. I\left(\frac{L}{t}\right)\right|_{t=T_0}$. Since $T_0$ satisfies $-\ln(N) = -T_0 I(\tfrac{L}{T_0})$ we can cancel terms and solve for $\delta$ in the resulting equation
$$
\begin{aligned}
0\approx& -\delta T_0 \partial_t\left. I\left(\tfrac{L}{t}\right)\right|_{t=T_0}-\delta I\left(\tfrac{L}{T_0}\right) - \delta^2 \partial_t\left. I\left(\tfrac{L}{t}\right)\right|_{t=T_0}+
 \\
&\ln\Big(e^{T_0^{1/3}\sigma\left(\frac{L}{T_0}\right)\chi_{L, T_0}} + e^{T_0^{1/3}\sigma\left(\frac{L}{T_0}\right)\chi_{-L, T_0}}\Big).
\end{aligned}
$$
Since $L$ and $T_0$ are both of order $\ln(N)$, it follows from the chain rule that $\partial_t\left. I\left(\tfrac{L}{t}\right)\right|_{t=T_0}$ is of order $(\ln(N))^{-1}$. Since the final term above is of order $(\ln(N))^{1/3}$ the only consistent scaling for $\delta$ is that it be of order $(\ln(N))^{1/3}$ as well in which case all terms are of that order except the term with $\delta^2$ which decays like $(\ln(N))^{-1/3}$. Thus, neglecting that term we solve for $\delta$ and find
\begin{equation}
\delta \approx \frac{\ln\Big(e^{T_0^{1/3}\sigma\left(\frac{L}{T_0}\right)\chi_{L, T_0}} + e^{T_0^{1/3}\sigma\left(\frac{L}{T_0}\right)\chi_{-L, T_0}}\Big)}{I\big(\frac{L}{T_0}\big) + T_0 \partial_t\left. I\big(\frac{L}{t}\big)\right|_{t=T_0}}.
\end{equation}

Therefore, we have shown that
\begin{equation}\label{eq:TWFPT}
 \env \approx T_0 + \frac{\ln\Big(e^{T_0^{1/3}\sigma\left(\frac{L}{T_0}\right)\chi_{L, T_0}} + e^{T_0^{1/3}\sigma\left(\frac{L}{T_0}\right)\chi_{-L, T_0}}\Big)}{I\big(\frac{L}{T_0}\big) + T_0 \partial_t\left. I\big(\frac{L}{t}\big)\right|_{t=T_0}}.
\end{equation}
The above conclusion is in agreement with Eq. 89 in \cite{doussalDiffusionTimedependentRandom2017a} (in the one-sided barrier case). In particular, our $\env$ random variable is essentially the same as $T_{\mathrm{Hit}}(\ell)$ with our $L$ and their $\ell$ having the same meaning. Our rate function $I$ is the same as their $\lambda$, and our $\hat{L}$ is equivalent to $1/\hat{\gamma}$ in their notation.

From Eq. \ref{eq:TWFPT} we may extract the following conclusion regarding the asymptotic behaviors for the mean and variance of $\env$ in the medium distance regime:
$$
\meanasy{\env} \approx M_1(L,N) \textrm{ and } \varasy{\env} \approx V_1(L, N)
$$
where $M_1$ and $V_1$ are defined as
\begin{equation}\label{eq:TWVariance}
\begin{aligned}
&M_1(L,N)\coloneqq \frac{(\ln(N))^2 + L^2}{2\ln(N)} \\
&V_1(L,N)\coloneqq
\frac{\mathrm{Var}\bigg(\!\!\ln\!\Big(e^{T_0^{1/3}\sigma\left(\frac{L}{T_0}\right)\chi_{L, T_0}} \!\!+ e^{T_0^{1/3}\sigma\left(\frac{L}{T_0}\right)\chi_{-L, T_0}}\Big)\!\!\bigg)}{\left(I\big(\frac{L}{T_0}\big) + T_0 \partial_t\left. I\big(\frac{L}{t}\big)\right|_{t=T_0}\right)^2}
\end{aligned}
\end{equation}
We have dropped the mean of $\delta$ (which is over order $(\ln(N))^{1/3}$) from $M_1$ and only retrained the first order term $T_0$. On the other hand, $V_1$ is precisely the variance of $\delta$ (as $T_0$ is deterministic).
 $V_1(L, N)$ contains the variance of a non-trivial combination of two random variables $\chi_{L, T_0}$ and $\chi_{-L, T_0}$. To estimate this variance we replace both by independent GUE TW random variables $\chi$ and $\chi'$ (as justified by the above recorded result of \cite{barraquandRandomwalkBetadistributedRandom2017}  and KPZ scaling theory. Under that replacement, the variance in the numerator in $V_1$ is
\begin{equation}
\begin{aligned}\label{eq:numeval}
& \int_{\mathbb{R}^2}dxdy \bigg(\ln\Big(e^{T_0^{1/3}\sigma\left(\frac{L}{T_0}\right) x} + e^{T_0^{1/3}\sigma\left(\frac{L}{T_0}\right)y} \Big)\bigg)^2 p(x) p(y) \\
&-\bigg( \int_{\mathbb{R}^2}dxdy\ln\Big(e^{T_0^{1/3}\sigma\left(\frac{L}{T_0}\right) x} + e^{T_0^{1/3}\sigma\left(\frac{L}{T_0}\right)y} \Big) p(x) p(y)\bigg)^2
\end{aligned}
\end{equation}
where $p(x)$ and $p(y)$ are the probability density of the GUE TW distribution. We numerically approximate the double integrals over the  $xy$ plane by integrating over the region $x,y\in [-10, 10]$ as is justified by the rapid decay of the density $p$. In fact, this integral can be approximated in the limit of large $N$ as follows. Observe that $T_0= \ln(N) \tfrac{\hat{L}^2+1}{2}$ where we let $L=\hat{L} \ln(N)$. Thus
$$
T_0^{1/3}\sigma\left(\tfrac{L}{T_0}\right) = (\ln(N))^{1/3} \left(\tfrac{\hat{L}^2+1}{2}\right) \sigma\left(\tfrac{2\hat{L}}{\hat{L}^2+1}\right).
$$
For fixed $\hat{L}$ this implies that the exponent diverges like $(\ln(N))^{1/3}$. Since
\begin{equation}\label{eq:rinf}
\ln(e^{rA}+e^{rB}) \approx r \max(A,B)\textrm{ as }r\to \infty,
\end{equation}
it follows that the numerator in Eq. \ref{eq:TWVariance} is approximately (as $N\to \infty$) given by
\begin{equation}\label{eq:rinfexp}
\Big(T_0^{1/3}\sigma\left(\tfrac{L}{T_0}\right) \Big)^2\var{\max(\chi,\chi')}
\end{equation}
where $\chi$ and $\chi'$ are independent GUE TW random variables. This variance can be computed via numerical integration and this need only be done once (as opposed to for various values of $N$ and $L$ as above).

\medskip
\noindent{\it One-sided case:} The same reasoning as in the two-sided case yields
expressions for the mean and variance of $\env$. Namely, $M_1$ and $V_1$ from Eq. \ref{eq:TWVariance} are replaced now by $\widetilde{M}_1$ and $\widetilde{V}_1$ where $M_1=\widetilde{M}_1$ and
$$
\widetilde{V}_1(L,N)\coloneqq \left(\frac{T_0^{1/3}\sigma\left(\frac{L}{T_0}\right)}{I\big(\frac{L}{T_0}\big) + T_0 \partial_t\left. I\big(\frac{L}{t}\big)\right|_{t=T_0}}\right)^2 \var{\chi}
$$
where $\chi$ is a GUE TW random variable such that $\var{\chi} \approx 0.813$. The simplification in $\widetilde{V}_1$ comes from the fact that only the $\chi_{L,T_0}$ term is present in the one-sided case -- thus rather than dealing with the variance of the $\ln$ of a sum of exponentials, the $\ln$ and exponential terms cancel and the simpler expression follows.

\subsection{Large Distance Regime $\env$ Behavior}
In the large distance regime we assume that $L,N\to \infty$ with $L/(\ln(N))^{3/2}\to \hathat{L}\in (0,\infty)$.
The key input from \cite{barraquandModerateDeviationsDiffusion2020} (see also Eq. (67) in the supplementary material in \cite{ledoussalDiffusionTimedependentRandom2017a}) is as follows:
For $v \in (0, \infty)$ and  $x=vt^{3/4}$
\begin{equation}\label{eq:ProbKPZ}
 \ln(\mathbb{P}^{\bold{B}}(X(t)\geq x)) \approx -\frac{x^2}{2t} - \frac{x^4}{12 t^3} + \ln\left(\frac{x}{t}\right) + h\left(0, \frac{x^4}{t^3}\right)
\end{equation}
where $h(y,s)$ denotes the random height at position $y$ and time $s$ of the narrow wedge solution to the Kardar-Parisi-Zhang (KPZ) equation
\begin{equation}\label{eq:KPZ}
\partial_s h(y, s) = \frac{1}{2}\partial_y^2 h(y,s) + \frac{1}{2}\left(\partial_y h(y, s)\right)^2 + \eta(y,s)
\end{equation}
where $\eta(y,s)$ is space-time white noise \cite{kardarDynamicScalingGrowing1986,sasamotoOneDimensionalKardarParisiZhangEquation2010,calabreseFreeenergyDistributionDirected2010,dotsenkoBetheAnsatzDerivation2010,amirProbabilityDistributionFree2011}. By symmetry Eq. \ref{eq:ProbKPZ} will also hold with $ \ln(\mathbb{P}^{\bold{B}}(X(t)\leq -x))$ and with an asymptotically independent fluctuation term $h'\left(0, \frac{x^4}{t^3}\right)$ that has the same law as $h$.

Similar to our analysis in the medium distance regime,
combining Eq. \ref{eq:UnboundedApprox} and Eq. \ref{eq:BC} yields
\begin{equation}
\begin{aligned}\label{eq:LongDistanceProb}
&\ln\big(\mathbb{P}^{\bold{B}}(\tauL \leq t)\big) \approx \\
& -\frac{L^2}{2t} - \frac{L^4}{12 t^3} + \ln\left(\frac{L}{t}\right) + \ln\Big(e^{h(0, \frac{L^4}{t^3})}+e^{h'(0, \frac{L^4}{t^3})}\Big)
\end{aligned}
\end{equation}
$\env$ is essentially $t$ such that $\mathbb{P}^{\bold{B}}(\tauL \leq t)=1/N$ which yields the implicit equation for $\env$
\begin{equation}
\begin{aligned}\label{eq:longexps}
-\ln(N) \approx& -\frac{L^2}{2\env} - \frac{L^4}{12 \left(\env\right)^3} + \ln\left(\frac{L}{\env}\right)+ \\
&\ln\Big(e^{h\big(0, \frac{L^4}{(\env)^3}\big)} + e^{h'\big(0, \frac{L^4}{(\env)^3}\big)}\Big)
\end{aligned}
\end{equation}
where $h$ and $h'$ are independent as  above. The first term on the right-hand side is dominant and solving $\ln(N)=\frac{L^2}{2T_0}$ yields the first order behavior of $\env\approx T_0$ with
$$
T_0 = \frac{L^2}{2\ln(N)}.
$$
Under our scaling of $L$, this term is of order $(\ln(N))^2$ while the other deterministic terms on the right-hand side of Eq. \ref{eq:longexps} are either order 1 or $\ln(\ln(N))$. Thus, for the sake of the mean of $\env$, $T_0$ will suffice. To study its variance we write $\env = T_0+\delta$ where $\delta \ll T_0$ contains the randomness of $\env$. Neglecting the lower order terms in Eq. \ref{eq:longexps} (the second and third terms on the right-hand side), Taylor expanding $-\frac{L^2}{2 \env} =-\frac{L^2}{2(T_0+\delta)} \approx -\frac{L^2}{2T_0} + \frac{L^2}{2T_0^2} \delta$, and substituting $T_0$ for $\env$ in the $h$ and $h'$ expressions (as is again justified by the KPZ scaling theory), we arrive at an approximation for
$$\delta\approx \frac{2 T_0^2}{L^2}\ln\Big(e^{h\big(0, \frac{L^4}{T_0^3}\big)} + e^{h'\big(0, \frac{L^4}{T_0^3}\big)}\Big).$$
From the above we may extract the following conclusion regarding the asymptotic behaviors for the mean and variance of $\env$ in the large distance regime:
$$
\meanasy{\env} \approx M_2(L,N) \textrm{ and } \varasy{\env} \approx V_2(L,N)
$$
where $M_1$ and $V_1$ are defined as
\begin{equation}\label{eq:TWVariance2}
\begin{aligned}
&M_2(L,N)\coloneqq \frac{L^2}{2\ln(N)} \\
&V_2(L,N)\coloneqq  \frac{L^4 \,\mathrm{Var}\bigg(\!\!\ln\Big(e^{h\big(0, \frac{8 (\ln(N))^3}{L^2}\big)} + e^{h'\big(0, \frac{8 (\ln(N))^3}{L^2}\big)}\Big)\!\!\bigg)}{4(\ln(N))^4} .
\end{aligned}
\end{equation}

Notice that for $L=(\ln(N))^{3/2}\hathat{L}$ (as in the large distance regime scaling) the KPZ equation time $\frac{8 (\ln(N))^3}{L^2}=8/\hathat{L}^2$. Thus, the calculation of the variance in $V_2$ requires numerically integration using the exact formula for the one-point distribution for the KPZ equation from \cite{sasamotoOneDimensionalKardarParisiZhangEquation2010,calabreseFreeenergyDistributionDirected2010,dotsenkoBetheAnsatzDerivation2010,amirProbabilityDistributionFree2011}.
This formula is non-trivial to compute numerically due to its complexity. However, \cite{prolhacPersonalCommunication2022} (used in the work of \cite{prolhacHeightDistributionKPZ2011}) contains the numerics for the density of $h(0,s)$ for $s\in \{0.25, 0.35, 0.5, 0.75, 1.2, 2, 3.5, 6.5, 13, 25, 50, 100, 250\}$. In the limit where $s\to 0$ or $s\to \infty$ the law of $h(0,s)$ converges to be Gaussian or GUE TW (with appropriate scaling) and thus we can combine these limiting behaviors with the numerical data to produce, via smooth interpolation, a curve
$$s\mapsto \mathrm{Var}\bigg(\!\!\ln\Big(e^{h(0,s)} + e^{h'(0, s)}\Big)\!\!\bigg)$$
for all $s$ that we use to numerically evaluate $V_2(L,N)$.

\medskip
\noindent{\it One-sided case:} The same reasoning as in the two-sided case yields expressions for the mean and variance of $\env$. Namely, $M_2$ and $V_2$ from  Eq. \ref{eq:TWVariance2} are replaced now by $\widetilde{M}_2$ and $\widetilde{V}_2$ where $\widetilde{M}_2=M_2$ and
$$
\widetilde{V}_2(L,N)\coloneqq  \frac{L^4 \,\mathrm{Var}\Big(h\big(0, \frac{8 (\ln(N))^3}{L^2}\big)\Big)}{4(\ln(N))^4} .
$$
In this case, the variance in question was numerically computed and plotted in \cite{prolhacHeightDistributionKPZ2011}.

\subsection{Stitching Together the Medium and Large Distance Regime $\env$ Behavior}

We compare the large $L$ (in the $\ln(N)$ scale) behavior of $V_1(L,N)$ to the small $L$ (in the $(\ln(N))^{3/2}$ scale) behavior of $V_2(L,N)$ and show that they match. This justifies defining $\varasy{\env}$ via smoothly stitching these two functions between these two scaling regimes. We then record the very large distance asymptotics that should persist for all $L$ beyond the $(\ln(N))^{3/2}$ regime. We only address the variances below since  $M_1(L,N)$ clearly converges to $M_2(L,N)$ when $L\gg \ln(N)$.

Recall $V_1(L,N)$ from Eq. \ref{eq:TWVariance}. Using the discussion after Eq. \ref{eq:numeval}, namely Eq. \ref{eq:rinfexp}, we can approximate this as
$$
V_1(L,N)\approx
\frac{\Big(T_0^{1/3}\sigma\big(\frac{L}{T_0}\big)\Big)^2 \mathrm{Var}\big(\max(\chi,\chi')\big)}{\left(I\big(\frac{L}{T_0}\big) + T_0 \partial_t\left. I\big(\frac{L}{t}\big)\right|_{t=T_0}\right)^2},
$$
where $\chi$ and $\chi'$ are independent GUE TW random variables.
Observe the following asymptotic: For $v\to 0$, $\sigma(v)\approx \frac{v^{4/3}}{2^{1/3}}$ while for $L\gg \ln(N)$, $T_0 \approx \frac{L^2}{2\ln(N)}$ and
$$
I\left(\frac{L}{T_0}\right) + T_0 \partial_t\left. I\left(\frac{L}{t}\right)\right|_{t=T_0}\approx -2\left(\frac{\ln(N)}{L}\right)^2.
$$
Putting these together shows that for $L\gg \ln(N)$,
$$
V_1(L,N)\approx
\frac{L^{8/3}\mathrm{Var}\big(\max(\chi,\chi')\big)}{2^{1/3} (\ln(N))^2}.
$$

Now, let us compare this to the behavior of $V_2(L,N)$ from Eq. \ref{eq:TWVariance2} when $L\ll (\ln(N))^{3/2}$.
Observe that the KPZ equation time $s$ in $V_2$ is given by $s=\frac{8(\ln(N))^3}{L^2}$ which goes to infinity as the ratio $L/(\ln(N))^{3/2}$ tends to zero. Thus, to extract the behavior of $V_2$ we must used the large time behavior of KPZ equation one-point distribution which says that the random variable $\chi_s$ defined by
$$
h(0, s) \approx \chi \left(\frac{s}{2}\right)^{1/3} - \frac{s}{24},
$$
converges as $s\to \infty$ to a GUE TW random variable \cite{sasamotoOneDimensionalKardarParisiZhangEquation2010, calabreseFreeenergyDistributionDirected2010, dotsenkoBetheAnsatzDerivation2010,amirProbabilityDistributionFree2011}. Letting $\chi$ and $\chi'$ denote the limiting independent GUE TW random variables arising from $h$ and $h'$ in $V_2$, and using Eq. \ref{eq:rinf}, it follows that when  $L\ll (\ln(N))^{3/2}$,
$$
V_2(L,N)\approx
\frac{L^{8/3}\mathrm{Var}\big(\max(\chi,\chi')\big)}{2^{1/3} (\ln(N))^2}
$$
which  matches the $L\gg \ln(N)$ behavior of $V_1(L,N)$.

This matching of the two expressions $V_1$ and $V_2$ justifies stitching them together to provide a single continuous curve for the asymptotic variance of the environmental fluctuations $\env$. To do this we use an error function centered at $L=(\ln(N))^{5/4}$ with a width of $(\ln(N))^{6/5}$ to ensure a smooth crossover between the two regimes. The resulting asymptotic variance formulas is then given by
\begin{equation}\label{eq:FullEnvVariance}
\varasy{\env} \!=\! \phi(L,N) V_1(L, N) + \big(1-\phi(L,N)\big) V_2(L, N)
\end{equation}
where $V_1(L, N)$ and $V_2(L, N)$ are defined in Eqs. \ref{eq:TWVariance} and \ref{eq:TWVariance2} respectively,
where the interpolation function
\begin{equation}\label{eq:interpo}
\phi(L,N)\coloneqq  \frac{1}{2} \left(1-\mathrm{erf}\left(\frac{L-(\ln(N))^{5/4}}{(\ln(N))^{6/5}}\right)\right)
\end{equation}
with the error function $\mathrm{erf}(x) = \frac{2}{\sqrt{\pi}}\int_0^x e^{-y^2}dy$. We likewise define $\meanasy{\env}$ via the same interpolation scheme as in Eq. \ref{eq:FullEnvVariance} using the fact that $M_1$ and $M_2$ smoothly crossover too. In fact, since the large $L$ asymptotic of $M_1$ is precisely in agreement with $M_2$, the interpolation is essentially unnecessary.

The above asymptotic formula $\varasy{\env}$ will be compared to numerical simulations for wide ranges of $N$ and $L$ in Section \ref{sec:compare}. As will become clear there, the crossover $L^{8/3}/(\ln(N))^2$ power-law behavior observed above is something of a ghost. For realistic sizes of $N$ the range  between $\ln(N)$ and $(\ln(N))^{3/2}$ is rather narrow. For instance, in Section \ref{sec:compare} we study the range $N=10^2$ to $N=10^{28}$. On the lower end of this range, $\ln(N)\approx 4.6$ and $(\ln(N))^{3/2}\approx 9.9$ while on the upper end $\ln(N)\approx 64$ while $(\ln(N))^{3/2}\approx 517$. So, even for $N=10^{28}$, there is not even a decade between $\ln(N)$ and $(\ln(N))^{3/2}$.

What is  more important in terms of comparison to numerical data is the behavior of the variance of $\env$ in the limit where $L\gg (\ln(N))^{3/2}$. This unbounded regime demonstrates a novel power-law that will be important to compare to the impact of the $\sam$ variance.

To probe the behavior of $\varasy{\env}$ as $L\gg (\ln(N))^{3/2}$ we need only study the corresponding behavior of $V_2(L,N)$. In that case the KPZ time $s=\frac{8(\ln(N))^3}{L^2}$ goes to zero. Thus, we must make use of the small time (Edwards-Wilkinson) asymptotics that show that the random variable $G_s$ defined by
\begin{equation}
h(0, s) \approx -\frac{s}{24} - \ln(\sqrt{2 \pi s}) + \left(\frac{\pi s}{4}\right)^{1/4} G_s
\end{equation}
converges as $s\to 0$ to a standard Gaussian random variable $G$ \cite{sasamotoOneDimensionalKardarParisiZhangEquation2010, calabreseFreeenergyDistributionDirected2010, dotsenkoBetheAnsatzDerivation2010,amirProbabilityDistributionFree2011}. Therefore, using $e^x \approx 1 + x$ and $\ln(1+x) \approx x$ as $x\to 0$, we find that
\begin{equation}
\begin{aligned}\label{eq:lneq}
\ln\!\big(e^{h(0,s)} \!+\! e^{h'(0,s)}\big)\approx -\tfrac{s}{24} - \ln\left(\!\sqrt{\tfrac{\pi s}{2}}\right) \!+\! \frac{1}{2}\left(\tfrac{\pi s}{4}\right)^{1/4}\!(G\! +\! G').
\end{aligned}
\end{equation}
Substituting $s = \frac{8(\ln(N))^3}{L^2}$ and taking the variance yields
$$
\mathrm{Var}\Big(\!\!\ln\Big(e^{h\big(0, \frac{8 (\ln(N))^3}{L^2}\big)} + e^{h'\big(0, \frac{8 (\ln(N))^3}{L^2}\big)}\Big)\!\!\Big) \approx \frac{1}{2}\! \left(\!\frac{2\pi(\ln(N))^3}{L^2}\right)^{1/2}\!\!\!,
$$
where we used that $\var{G+G'}=2$ since $G$ and $G'$ are independent standard Gaussian random variables. Substituting this into Eq. \ref{eq:TWVariance2}
shows that where $L\gg (\ln(N))^{3/2}$
\begin{equation}\label{eq:KPZVar}
\varasy{\env}\approx V_2(L,N)\approx   \frac{1}{4}\sqrt{\frac{\pi}{2}}\frac{L^{3}}{(\ln(N))^{5/2}}.
\end{equation}
This power-law behavior will be quite visible in the numerical data from Section \ref{sec:compare}.

\medskip
\noindent{\it One-sided case:} The only difference in this case is that we use $\widetilde{V}_1$ and $\widetilde{V}_2$ in place of $V_1$ and $V_2$. This crossover behavior between the regime where $L$ is of order $\ln(N)$ and $(\ln(N))^{3/2}$ still matches up and the same interpolation formula Eq. \ref{eq:FullEnvVariance} can be used. For the $L\gg (\ln(N))^{3/2}$ asymptotics, the right-hand side of Eq. \ref{eq:KPZVar} ends up being twice as large in the one-sided case as in the two-sided case. This may seem counter-intuitive since there are two Gaussians in the two-sided case versus one in the one-sided case. However, in the calculation in Eq. \ref{eq:lneq}, we used
$\ln\big(e^{h(0,s)} + e^{h'(0,s)}\big) \approx \ln(2) + \frac{1}{2}\big(h(0,s) +h'(0,s)\big)$. The variance of $\frac{1}{2}\big(h(0,s) +h'(0,s)\big)$ is half that of the variance of $h(0,s)$, thus explaining the factor of $2$ difference.

\subsection{$\sam$ and $\min$ Behavior}

We now argue that the environmental and sampling fluctuations $\env$ and $\sam$ are independent and that the sampling fluctuations are Gumbel distributed. The independence is strongly supported by our numerical data, see for instance Figure \ref{fig:EnvRecovery}.

To start, observe that for large $N$ and small $\mathbb{P}^{\bold{B}}(\tauL \leq t)$ we can rewrite
Eq. \ref{eq:NFPT} as
\begin{equation}\label{eq:NFPTApprox}
\mathbb{P}^{\bold{B}}(\min \geq t) \approx  e^{-N \mathbb{P}^{\bold{B}}(\tauL \leq t)}.
\end{equation}
Using the non-backtracking approximation in Eq. \ref{eq:UnboundedApprox} and the medium distance range asymptotic expansion Eq. \ref{eq:BC} we arrive at the starting formula
\begin{equation}
\begin{aligned}\label{eq:lnPsum}
&\ln\big(\mathbb{P}^{\bold{B}}(\min \geq t)\big) \approx \\
&-N\Big(
e^{- t I\left(\tfrac{L}{t}\right)+t^{1/3}\sigma(\frac{L}{t})\chi_{L, t}} + e^{- t I\left(\tfrac{L}{t}\right)+t^{1/3}\sigma(\frac{L}{t})\chi_{-L, t}}\Big)
\end{aligned}
\end{equation}
We will work here with in the medium distance regime asymptotics. A similar analysis could be performed in the large distance regime and the result would agree with the $L\gg \ln(N)$ behavior derived below so we do not repeat this derivation.

Recalling that $\sam=\min-\env$, we have that $\mathbb{P}^{\bold{B}}(\sam \geq s) = \mathbb{P}^{\bold{B}}(\min \geq \env+s)$. Thus we have that $\ln\big(\mathbb{P}^{\bold{B}}(\sam \geq s)\big)$ is approximately given by the right-hand side of Eq. \ref{eq:lnPsum} with $t=\env+s$. With this in mind, we will use the following Taylor expansion in the exponential on the right-hand side of Eq. \ref{eq:lnPsum}:
\begin{equation}
\begin{aligned}
&- (\env+s) I\left(\tfrac{L}{\env+s}\right)\\
&\qquad+(\env+s)^{1/3}\sigma\big(\tfrac{L}{\env+s}\big)\chi_{L,\env+s} \approx \\
&- (\env+s) \Big(I\left(\tfrac{L}{\env}\right)+s\partial_t \left. I\left(\tfrac{L}{t}\right)\right|_{t=\env}\Big) \\ &\qquad+(\env)^{1/3}\sigma\big(\tfrac{L}{\env}\big)\chi_{L, \env} \approx \\
&\ln\big(\mathbb{P}^{\bold{B}}(X(\env)  \geq L)\big) -\\
&\qquad s \Big(I\left(\tfrac{L}{\env}\right)+ \env \partial_t \left. I\left(\tfrac{L}{t}\right)\right|_{t=\env}\Big)
\end{aligned}
\end{equation}
Notice that in the first comparison above we have assumed that $(\env+s)^{1/3}\sigma\big(\tfrac{L}{\env+s}\big)\chi_{L,\env+s}$ can be replaced by the same term with $\env$ instead of $\env+s$. This is a non-trivial assumption which ultimately implies the Gumbel form of $\sam$ and its independence from $\env$. This should be justifiable based on the KPZ scaling theory and the fact that (as we will see) $\meanasy{\env}\gg \meanasy{\sam}$.  The second comparison uses Eq. \ref{eq:BC} and throws out the $s^2$ term. Of course, the same expansion above applies when $L\mapsto -L$ and $\mathbb{P}^{\bold{B}}(X(\env)  \geq L)$ is replaced by $\mathbb{P}^{\bold{B}}(X(\env) \leq -L)$. Putting these expansions together with Eq. \ref{eq:lnPsum} yields
\begin{equation}
\begin{aligned}\label{eq:lnpbnolz}
&\ln\big(\mathbb{P}^{\bold{B}}(\sam \geq s)\big) \approx \\
&-N\big(\mathbb{P}^{\bold{B}}(X(\env)  \geq L)+\mathbb{P}^{\bold{B}}(X(\env)  \leq -L)\big) \\
&\times e^{- s \big(I\left(\tfrac{L}{\env}\right)+\env \partial_t \left. I\left(\tfrac{L}{t}\right)\right|_{t=\env}\big)}
\end{aligned}
\end{equation}
Invoking the non-backtracking approximation in Eq. \ref{eq:UnboundedApprox} and the definition Eq. \ref{eq:EnvDef} of $\env$, it follows that $\mathbb{P}^{\bold{B}}(X(\env)  \geq L)+\mathbb{P}^{\bold{B}}(X(\env)  \leq -L) \approx \mathbb{P}^{\bold{B}}(\tauL \leq \env)\approx 1/N$. Using this along with replacing $\env$ by $T_0$ as in Eq. \ref{eq:TWFirstOrder}, from Eq. \ref{eq:lnpbnolz} we see that
$$
\ln\big(\mathbb{P}^{\bold{B}}(\sam \geq s)\big) \approx
-e^{- s \big(I\left(\tfrac{L}{T_0}\right)+T_0 \partial_t \left. I\left(\tfrac{L}{t}\right)\right|_{t=T_0}\big)}.
$$
Observe that
\begin{equation*}
\begin{aligned}
I\left(\tfrac{L}{T_0}\right)+T_0 \partial_t \left. I\left(\tfrac{L}{t}\right)\right|_{t=T_0} &= I(v)-v I'(v)\Big\vert_{v=L/T_0}\\
&=\frac{\sqrt{1-v^2}-1}{\sqrt{1-v^2}}\Big\vert_{v=L/T_0},
\end{aligned}
\end{equation*}
which is negative and behaves like $-v^2/2$ as $v\to 0$. This shows that asymptotically $-\sam$ has the law of a Gumbel distribution with location parameter $0$ and scale parameter $-\Big(I\left(\tfrac{L}{T_0}\right)+T_0 \partial_t \left. I\left(\tfrac{L}{t}\right)\right|_{t=T_0}\Big)^{-1}$. From this and the formula for the mean and variance of a Gumbel random variable in terms of its location and scale parameter we see that in the medium distance regime
\begin{equation}\label{eq:RWRESam}
\begin{aligned}
 \meanasy{\sam} &\approx \frac{\gamma}{\left(I\left(\frac{L}{T_0}\right) + T_0 \partial_t \left. I\left(\frac{L}{t}\right)\right|_{t=T_0}\right)}, \\
 \varasy{\sam} &\approx \frac{\pi^2}{6 \left(I\left(\frac{L}{T_0}\right) + T_0 \partial_t \left. I\left(\frac{L}{t}\right)\right|_{t=T_0}\right)^2},
 \end{aligned}
\end{equation}
where $\gamma\approx .577$ is the Euler gamma constant.
In the limit where $L\gg \ln(N)$ this yields
\begin{equation}\label{eq:SamVar}
\begin{aligned}
 \meanasy{\sam} &\approx -\frac{\gamma L^2}{2(\ln(N))^2},\\
 \varasy{\sam} &\approx \frac{\pi^2}{24} \frac{L^4}{(\ln(N))^4}.
\end{aligned}
\end{equation}

\begin{figure}[t]
 \centering
 \includegraphics[width=\columnwidth]{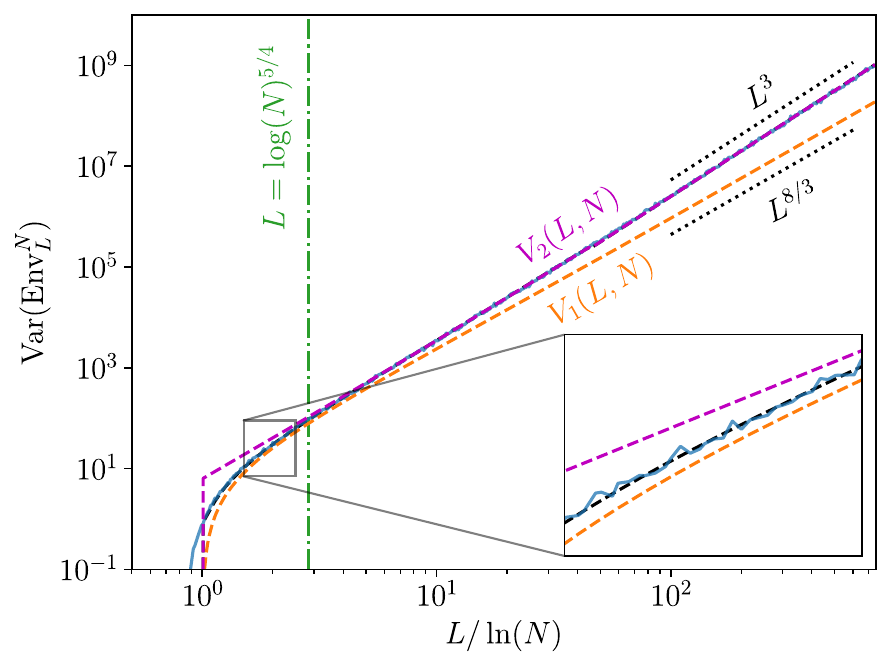}
 \caption{We plot the environmental variance, $\varnum{\env}$, for $N=10^{28}$ particles (blue); the asymptotic variance $V_1$ for the short time regime given in Eq. \ref{eq:TWVariance} (orange dashed line); the asymptotic variance $V_2$ for the long time regime given in Eq.  \ref{eq:TWVariance2} (purple dashed line); the interpolation $\varasy{\env}$ between these regimes using Eq. \ref{eq:FullEnvVariance} (black dashed line); and the power-law asymptotics of $V_1$ and $V_2$ (black dotted lines).}
 \label{fig:VarianceStitching}
\end{figure}

Observe that $\meanasy{\sam} \ll \meanasy{\env}$ since the former grows like $L^2/(\ln(N))^2$ while the latter like $L^2/\ln(N)$. Thus, we are justified in dropping $\meanasy{\sam}$ and approximating $\meanasy{\min} \approx \meanasy{\env}$ as we will do in Eq. \ref{eq:meanasymin}.

Though the above derivation was done in the medium distance regime, the same could be repeated in the large distance regime and the above asymptotic behavior in Eq. \ref{eq:SamVar} would follow (along with the Gumbel scaling limit for $\sam$. We do not repeat this calculation here. This, however, justifies using the asymptotic formula from Eq. \ref{eq:RWRESam} for both the medium and large distance regimes (and $L\gg (\ln(N))^{3/2}$ too).

From the above explained Gumbel limit, observe that the location and scale parameters are independent of $\env$ and rather only depend on the first order behavior of $\env$ given by $T_0$. This implies the asymptotic independence and hence allows us to represent $\min$ as a sum of $\env$ (whose limiting distribution, mean and variance were identified earlier) and $\sam$ (which is Gumbel distributed as above). This independence implies the following {\it addition law}
\begin{equation}\label{eq:addlaw}
\varasy{\min}=\varasy{\env}+\varasy{\sam}.
\end{equation}
The same trivially holds for the mean.

Therefore, combining Eq. \ref{eq:FullEnvVariance} with Eq. \ref{eq:RWRESam} we conclude (see also Figure \ref{fig:VarianceStitching})
\begin{equation}
\begin{aligned}\label{eq:varasymin}
 \varasy{\min} \approx& \phi(L,N) V_1(L, N)\! +\! \big(1-\phi(L,N)\big) V_2(L, N) \\
 &+ \frac{\pi^2}{6 \left(I\left(\frac{L}{T_0}\right) + T_0 \partial_t \left. I\left(\frac{L}{t}\right)\right|_{t=T_0}\right)^2}
 \end{aligned}
\end{equation}
where $V_1(L, N)$ and $V_2(L, N)$ are defined in Eqs. \ref{eq:TWVariance} and \ref{eq:TWVariance2} respectively, the interpolation function $\phi$ is defined in Eq. \ref{eq:interpo} and the final term comes from Eq. \ref{eq:RWRESam}. Similarly,
\begin{equation}
\begin{aligned}\label{eq:meanasymin}
 \meanasy{\min} \approx& M_1(L, N).
 \end{aligned}
\end{equation}

It is worth emphasizing (and will be apparent in the numerical simulation data that follows) that in the asymptotic formula $\varasy{\min}$ in Eq. \ref{eq:varasymin}, there is a competition between $\varasy{\env}$ and $\varasy{\sam}$. By Eq. \ref{eq:KPZVar} $\varasy{\env}$ behaves (anywhere past the short lived medium distance regime) according to the power-law $L^3/(\ln(N))^{5/2}$ while by Eq. \ref{eq:SamVar} $\varasy{\sam}$ likewise behaves like $L^4/(\ln(N))^{3/2}$. Setting these equal shows a crossover when $L$ is of order $(\ln(N))^{3/2}$, i.e., the large distance regime. For smaller $L$, $\varasy{\env}\gg \varasy{\sam}$ while for larger $L$ the opposite holds.

\medskip
\noindent{\it One-sided case:} The behavior of $\sam$ is easily seen to be asymptotically the same in this case. The behavior of $\min$ thus involves combining the one-sided behavior of $\env$ described earlier with the above behavior of $\sam$.

\subsection{Asymptotic Behavior for the SSRW}

We now derive the extreme first passage time behavior for the SSRW model where the transition biases are deterministic with $B(x,t)=1/2$. We derive these results using the same techniques and assumptions used for the \rwre{} model. The methods used in \cite{rednerGuideFirstPassageProcesses2001a, lawleyDistributionExtremeFirst2020} should also yield an alternative derivation of the asymptotics below though we do not pursue that here. To distinguish from the \rwre{} model, we will use a tilde to label random variables and functions associated to the SSRW model below.

Just as for the \rwre{} model, we find a trivial short distance behavior for the SSRW model but with a larger cutoff on the short distance length scale. Namely, when $L,N\to\infty$ with $L/\ln(N)\to \hat{L}<1/\ln(2)$ the mean and variance of the extreme first passage time behaves asymptotically like $\meanasy{\minSSRW} \approx L$ and $\varasy{\minSSRW}\approx 0$. The $1/\ln(2)$ comes from solving for $L$ such that $(1/2)^L\approx 1/N$ (we do not need to use the law of large numbers here since all $B(x,t)=1/2$).

The other length regime is when $L,N\to\infty$ with $L/\ln(N)\to \hat{L}>1/\ln(2)$ (or when $L/\ln(N)$ goes to infinity).
Using Stirling's formula (or more generally Cramer's theorem from the large deviation theory) the tail of the probability distribution for the location of a SSRW $\tilde{X}(t)$ satisfies
\begin{equation}\label{eq:SSRWDistribution}
 \mathbb{P}(\tilde{X}(t) \geq x) \approx e^{-t\ISSRW\left(\frac{x}{t}\right)}
\end{equation}
where $\ISSRW(v) = \frac{1}{2}\left((1+v)\ln(1+v) + (1-v)\ln(1-v)\right)$ and where we no longer write $\mathbb{P}^{\bold{B}}$ since deterministically we have assumed here that all $B(x,t)=1/2$.

Using the same non-backtracking approximation as in the \rwre{} case (Eq. \ref{eq:UnboundedApprox}) in conjunction with Eq. \ref{eq:SSRWDistribution} yields
\begin{equation}\label{eq:SSRWFPT}
\mathbb{P}(\widetilde{\tauL} \leq t) \approx 2e^{-t\ISSRW(\frac{L}{t})}.
\end{equation}

We look for $\tilde{T}_0$ such that $\mathbb{P}(\widetilde{\tauL} \leq \tilde{T}_0) = 1/N$. This is a deterministic analog to $\env$ and plays the role of the centering of the extreme first passage time. Since $\tilde{T}_0$ is deterministic for the SSRW, there are no environmental fluctuations. Dropping the factor of $2$ in Eq. \ref{eq:SSRWFPT} (as it is insignificant in the asymptotic regimes we consider here) $\tilde{T}_0$ should satisfy
\begin{equation}\label{eq:SSRWFirstOrder}
\ln\left(N\right) \approx \tilde{T}_0 \, \cdot \, \ISSRW\left(\frac{L}{\tilde{T}_0}\right).
\end{equation}
Although we cannot solve this analytically for $\tilde{T}_0$, we can solve for $\tilde{T}_0$ numerically and asymptotically in the limit $\tilde{T}_0\rightarrow\infty$ (which occurs when $L$ grows fast enough compared to $\ln(N)$) as we show below.

Substituting Eq. \ref{eq:SSRWFPT} into Eq. \ref{eq:NFPTApprox} and expanding about $\tilde{T}_0$ such that $\minSSRW = \tilde{T}_0 + s$ yields
$$
\mathbb{P}(\minSSRW \geq \tilde{T}_0 + s) \approx e^{-2Ne^{-(\tilde{T}_0 + s)\ISSRW\left(\frac{L}{\tilde{T}_0 + s}\right)}}.
$$

\begin{figure}[t]
 \includegraphics[width=.98\columnwidth]{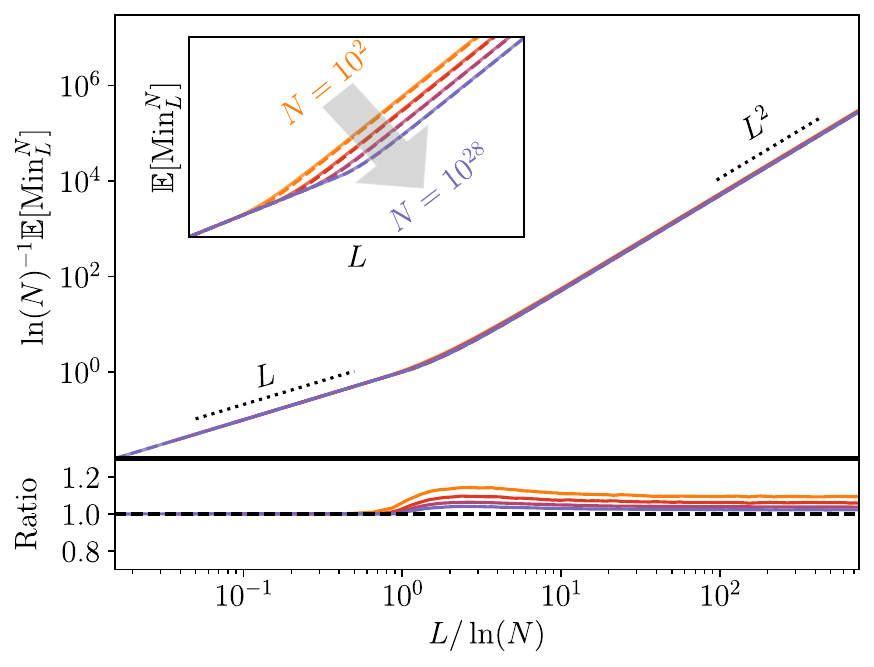}
 \caption{We plot the mean of the extreme first passage time, $\meannum{\min}$, using solid lines of varying colors for $N=10^2, 10^5, 10^{12}$ and $10^{28}$ particles. The dashed lines are the asymptotic curves $\meanasy{\min}$ from Eq. \ref{eq:meanasymin}. The inset shows the uncollapsed data to better distinguish different values of $N$. The collapse occurs by scaling both the $x$-axis and $y$-axis by $\ln(N)$. The lower plot shows the ratio of $\meannum{\min} / \meanasy{\min}$ confirming the close fit.}
 \label{fig:MaxMean}
\end{figure}

Expanding about $\tilde{T}_0$ such that $\ISSRW\left(\frac{L}{\tilde{T}_0 + s}\right)\approx \ISSRW\left(\frac{L}{\tilde{T}_0}\right)+s \partial_t \left. \ISSRW\left(\frac{L}{t}\right)\right|_{t=\tilde{T}_0}$ gives
$$
\mathbb{P}(\minSSRW \geq \tilde{T}_0 + s) \approx e^{-e^{-s \left(\ISSRW\left(\frac{L}{\tilde{T}_0}\right) + \tilde{T}_0 \partial_t \left.\ISSRW\left(\frac{L}{t}\right)\right|_{t=\tilde{T}_0}\right)}}
$$
This shows $-\minSSRW$ is Gumbel distributed with location parameter $-\tilde{T}_0$ and scale parameter $-\left(\ISSRW\left(\frac{L}{\tilde{T}_0}\right) + \tilde{T}_0 \partial_t \left.\ISSRW\left(\frac{L}{t}\right)\right|_{t=\tilde{T}_0}\right)^{-1}$ (which can be checked to be positive as needed to define a Gumbel distribution). Combining the above calculations we conclude that for the SSRW
\begin{equation}
\begin{aligned}\label{eq:SSRWVar}
\meanasy{\minSSRW} &\approx \tilde{T}_0, \\
 \varasy{\minSSRW} &\approx \frac{\pi^2}{6 \left(\ISSRW\left(\frac{L}{\tilde{T}_0}\right) + \tilde{T}_0 \partial_t \left. \ISSRW\left(\frac{L}{t}\right)\right|_{t=\tilde{T}_0}\right)^2}.
\end{aligned}
\end{equation}
Notice that we have dropped the lower order term $\gamma \left(\ISSRW\left(\frac{L}{\tilde{T}_0}\right) + \tilde{T}_0 \partial_t \left.\ISSRW\left(\frac{L}{t}\right)\right|_{t=\tilde{T}_0}\right)^{-1}$ from the approximation to $\meanasy{\minSSRW}$ above, just as we did in the \rwre{} case.
In the limit $L\gg \ln(N)$ we also have $\tilde{T}_0 \gg L$. Thus, using the expansion $\ISSRW(v)\approx v^2/2$ as $v\to 0$ we can solve for $\tilde{T}_0$ and thus extract the following asymptotics, valid in the $L\gg \ln(N)$  limit:
\begin{equation}\label{eq:SSRWVar}
\begin{aligned}
 \meanasy{\minSSRW} &\approx \frac{L^2}{2\ln(N)}, \\
 \varasy{\minSSRW} &\approx \frac{\pi^2}{24}\frac{L^4}{(\ln(N))^4}.
\end{aligned}
\end{equation}
These match the corresponding asymptotic sampling mean and variance formulas for the \rwre{} in Eq. \ref{eq:SamVar}.

\medskip
\noindent{\it One-sided case:} The same argument and result follows.
\begin{figure}[t]
 \includegraphics[width=\columnwidth]{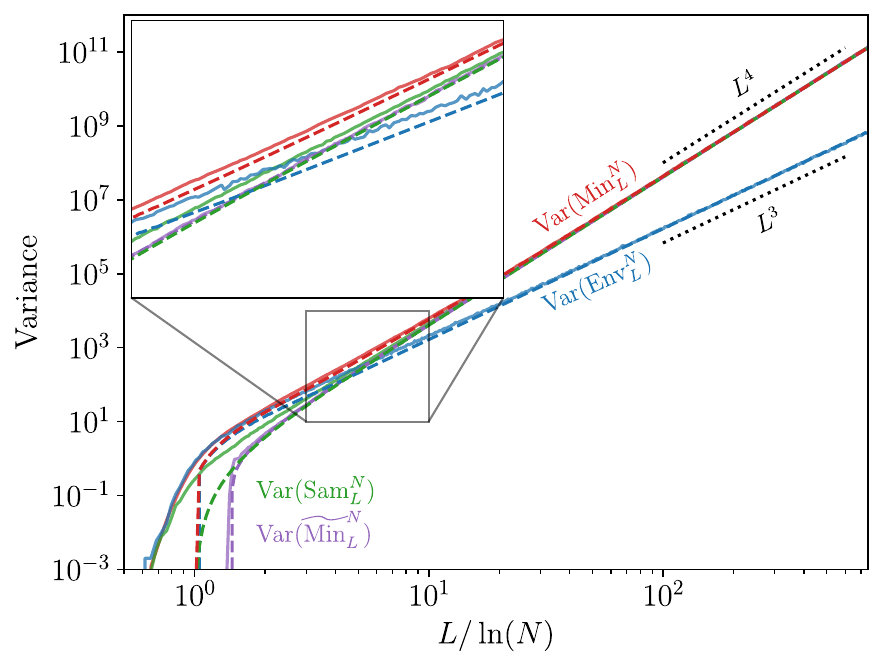}
 \caption{For the \rwre{} model with $N=10^{12}$ particles, we plot we plot the numerically measured (solid) and asymptotic theory (dashed) variance of the extreme first passage time $\var{\min}$ (red); the variance due to the environment $\var{\env}$ (blue); the variance due to sampling $\var{\sam}$ (green). For the SSRW model we plot $\var{\minSSRW}$ (purple).}
 \label{fig:CompleteVariance}
\end{figure}

\section{Comparison of Asymptotic and Numerical Results}\label{sec:compare}

\begin{figure}[t]
 \includegraphics[width=\columnwidth]{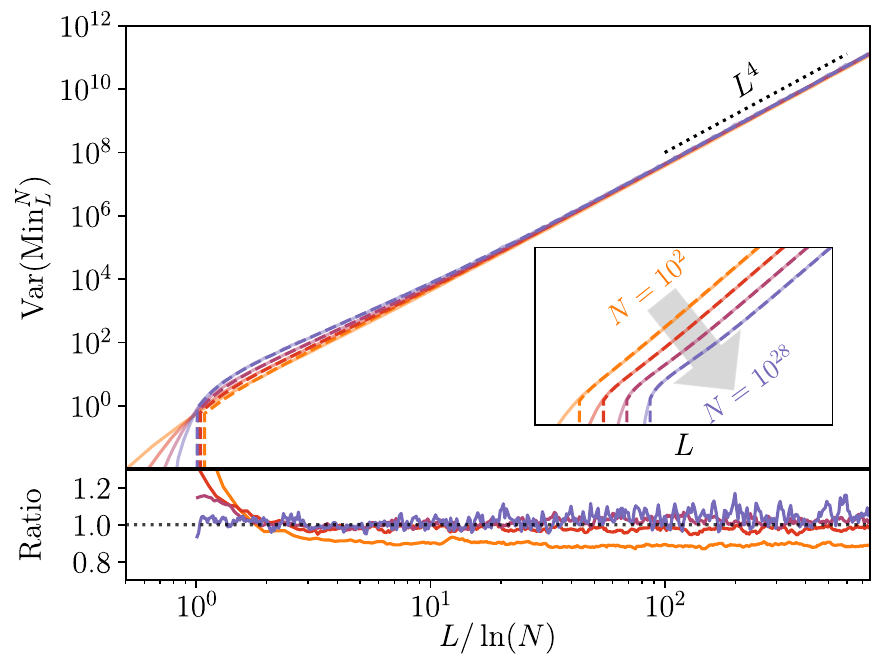}
 \caption{We plot the numerically measures (solid) and asymptotic theory (dashed) variance of extreme first passage time, $\var{\min}$, for $N=10^2, 10^5, 10^{12}$ and $10^{28}$ (each labeled with a different color). The asymptotic theory variance, $\varasy{\min}$, comes from Eq. \ref{eq:varasymin}. The inset shows the same data uncollapsed to better distinguish different $N$. The lower plot shows the ratio $\varnum{\min}/\varasy{\min}$.}
 \label{fig:MaxVariance}
\end{figure}

\begin{figure}[t]
 \includegraphics[width=\columnwidth]{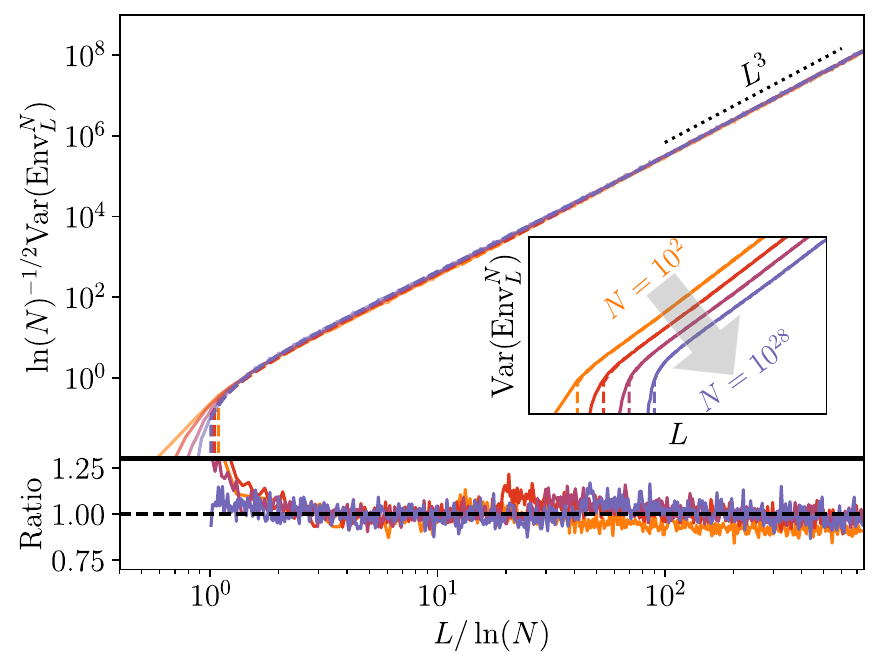}
 \caption{We plot the numerically measures (solid) and asymptotic theory (dashed) environmental variance, $\var{\env}$, for $N=10^2, 10^5, 10^{12}$ and $10^{28}$ (each labeled with a different color). The asymptotic theory variance, $\varasy{\env}$, comes from Eq. \ref{eq:FullEnvVariance}. The inset shows the same data uncollapsed to better distinguish different $N$. The lower plot shows the ratio $\varnum{\env}/\varasy{\env}$.}
 \label{fig:Environmental Variance}
\end{figure}

\begin{figure}[t]
\includegraphics[width=\columnwidth]{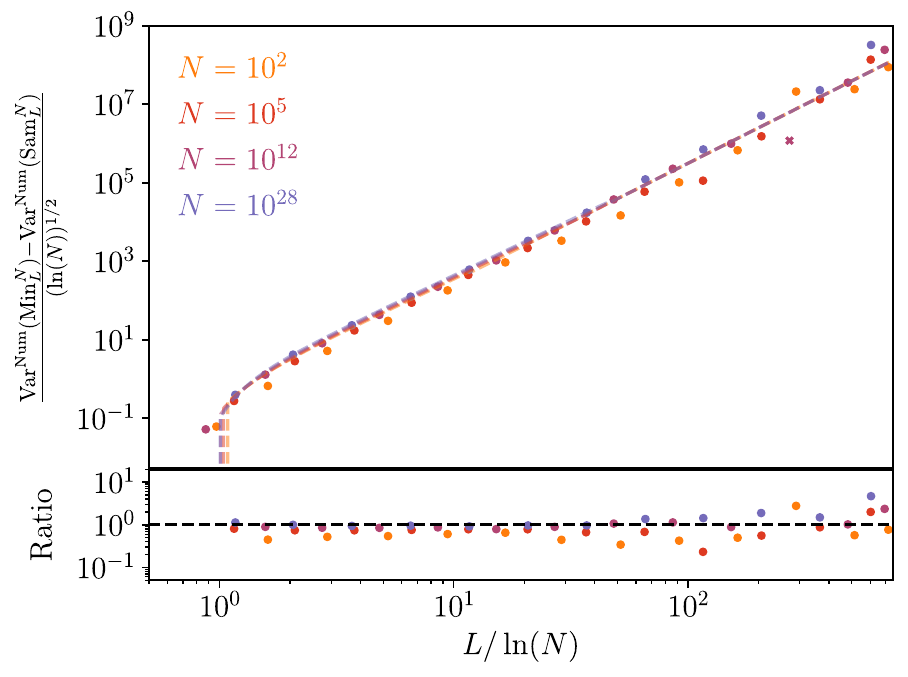}
\caption{We plot $\varnum{\min}-\varnum{\sam}$ for $N=10^2, 10^5, 10^{12}$ and $10^{28}$ averaged over each horizontal $1/4$ decade. The dashed line is $\varasy{\env}$ in Eq. \ref{eq:FullEnvVariance} for each $N$ corresponding to its respective color.}
\label{fig:EnvRecovery}
\end{figure}

We find excellent agreement between our asymptotic theory and numerical simulations not only for very large $N$, but even for as few as $N=100$ particles. These results are summarized through a number of figures. As a convention we use solid curves (or data points in Figures \ref{fig:EnvRecovery} and \ref{fig:EnvRecoveryAsym}) to record outcomes of numerical simulations, dashed curves to record our asymptotic theory, and dotted lines to denote relevant power-laws. All plots are in $\log$-$\log$ coordinates so power-laws are straight lines. In Figures \ref{fig:MaxMean},\ref{fig:MaxVariance}, \ref{fig:Environmental Variance},  \ref{fig:EnvRecovery} and \ref{fig:EnvRecoveryAsym} each color corresponds to a different value of $N$ while in Figure \ref{fig:CompleteVariance} a single value of $N$ is taken and the colors correspond to numerical and asymptotic curves for the variance of $\min,\env$ or $\sam$.

Figure \ref{fig:VarianceStitching} shows how we stitch together in Eq. \ref{eq:FullEnvVariance} the medium and large distance regimes to produce $\varasy{\env}$. It shows that this interpolation scheme provides a smooth crossover between the two regimes and agrees with numerical measurements.

Figure \ref{fig:MaxMean} shows the comparison of the numerical measurements and asymptotic theory for the mean of the extreme first passage time $\min$ for various $N$. For system sizes ranging from $N=10^2$ to $10^{28}$ the data and asymptotic curves are nearly indistinguishable and fall onto the same master curve given by Eq. \ref{eq:meanasymin}. The ratio of $\meannum{\min} / \meanasy{\min}$ is nearly $1$ for every $N$ indicating the numerics and asymptotic predictions are in agreement. The region with power-law $L$ corresponds to the short distance ballistic regime while the $L^2$ power-law is valid for all times from the medium distance regime on.

Figure \ref{fig:CompleteVariance} shows the numerical measurements and asymptotic theory for the variance of $\min, \sam$ and $\env$. The theory closely matches the numerical results except for $L$ very close to $\ln(N)$. Presumably, this is due to the finite system size and could be partially remedied by studying the crossover from the short to medium distance regime, for instance using the central limit theorem. In the medium distance regime, the numerical data $\varnum{\min}$ closely matches  $\varnum{\env}$ as well as the asymptotic formula $\varasy{\env}$. In the large distance regime,  $\varnum{\min}$ now closely matches  $\varnum{\sam}$ as well as the asymptotic formula $\varasy{\sam}$. This is because in the medium distance regime  $\var{\sam} \ll \var{\env}$ while in the large distance regime  $\var{\sam} \gg \var{\env}$. The interpolation curve $\varasy{\min}$ closely matches $\varnum{\min}$ over the full medium and long distance regime. For the SSRW,  $\varnum{\minSSRW}$ closely agrees with $\varasy{\minSSRW}$ (which, as noted earlier, is quite close to $\varasy{\sam}$ from the \rwre{} model for $L$ beyond the medium distance regime.

Figure \ref{fig:MaxVariance} and \ref{fig:Environmental Variance} show the numerically measured variances of $\min$ and $\env$ for a range of  system sizes as well as the corresponding asymptotic theory results given by Eqns. \ref{eq:varasymin} and \ref{eq:FullEnvVariance}, respectively. For large distances, the asymptotic predictions and numerical results match for all system sizes, ranging from $N=10^2$ to $10^{28}$. For $L/\ln(N)\approx 1$ we see additional variance due to finite size effects blunting the transition between the ballistic and diffusive regimes. The ratio of numerics to asymptotic predictions for $\min$ and $\env$ are nearly $1$ for large $L/\ln(N)$ indicating a good agreement between the numerics and asymptotic predictions. The asymptotic ratio approaches $1$ for larger $N$. However, even for $N=10^2$ the asymptotic ratio is about $.9$.

The close agreement between $\varnum{\min}$ and $\varasy{\min}$ in Figure \ref{fig:MaxVariance} provides a verification of the theoretical addition law, Eq. \ref{eq:addlaw}. Figures \ref{fig:EnvRecovery} and \ref{fig:EnvRecoveryAsym}  provide further confirmation of the addition law. In particular, Figure \ref{fig:EnvRecovery} shows that the numerically measured difference $\varnum{\min}-\varnum{\sam}$ follows the asymptotic theory curve $\varasy{\env}$. The power-law is clearly recovered, though there is considerably more variation in the ratio of $(\varnum{\min}-\varnum{\sam})/\varasy{\env}$ than in similar ratios observed in previous figures. This is not so surprising since the difference of two numerical measures introduces additional error. Also for larger $N$, the number of systems used to estimate the variances is smaller than for small $N$, thus introducing additional error for large $N$.



Figure \ref{fig:EnvRecoveryAsym} shows another route to measure the environmental variance $\var{\env}$ by taking the difference $\varnum{\min}-\varasy{\minSSRW}$. As we explain in Section \ref{sec:con}, we use $\varasy{\minSSRW}$ instead of $\varasy{\sam}$ here since in experimental settings it may be possible to estimate the Einstein diffusion coefficient and hence develop a prediction for $\varasy{\minSSRW}$. Since we have shown above that $\varasy{\minSSRW}$ and $\varasy{\sam}$ match beyond the medium distance regime, we thus expect to still be able to recover the power-law behavior of $\var{\env}$ by studying $\varnum{\min}-\varasy{\minSSRW}$.

In Figure \ref{fig:EnvRecoveryAsym}, the dots record positive values of this difference $\varnum{\min}-\varasy{\minSSRW}$ (scaled by $(\ln(N))^{1/2}$, as one should to see a collapse of the data) and the $\times$ records the magnitude of negative values of this difference. Clearly, the presence of negative values contradicts the addition law and recovery of $\var{\env}$. These values, however, only arise for either small $N$ ($10^2$ and $10^5$) or large $L$. For large $N$ ($10^{12}$ and $10^{28}$) the dots closely follow the asymptotic environmental variance curve $\varasy{\env}$ for just under two decades (as evidenced from the ratio being close to 1) and then peel off following what looks to be an $L^4$ power-law (as opposed to the $L^3$ power-law behavior of $\varasy{\env}$). The explanation for the lack of agreement for small $N$ or large $L$ likely arises from the presence of higher order corrections to $\varasy{\minSSRW}$ which scale like $L^4$ but with pre-factors that decay with $N$ (the presence of such terms can be seen from \cite{zarfatyAccuratelyApproximatingExtreme2021a}). Therefore, for small $N$ or large $L$ these corrections become relevant.

\section{Conclusion}\label{sec:con}
We consider two models for many-particle diffusion in a common environment. The first treats each particle as an independent SSRW while the second \rwre{} model treats the environment as a space-time random biasing field within which each particle performs biases random walks. We focus on the extreme first passage time $\min$, i.e., the time when the first of $N$ particles passes a barrier distance $L$ From their common starting location. We show that the randomness of $\min$ splits into two essentially independent pieces, the randomness $\env$ from the environment and the randomness $\sam$ from sampling $N$ random walks within that environment.

We determine theoretical predictions (related to the KPZ universality class and equation) for the behavior of each of these contributions based on asymptotic limit theorems in different scaling regimes of $N$ and $L$.
While $\var{\sam}$ closely matches the variance from sampling in the SSRW model for large $L$, the $\var{\env}$ term has no parallel in the SSRW case where the environment is deterministic. We uncover a novel $L^3/(\ln(N))^{5/2}$ power-law describing the large $L$ behavior of $\var{\env}$. This should be contrasted with the $L^4/(\ln(N))^4$ power-law that we demonstrate describes the large $L$ behavior of $\var{\sam}$. Thus, for large $L$, owing to the independence of $\env$ and $\sam$, we see that $\var{\min}=\var{\env}+\var{\sam}$ and only the $L^4$ power-law is visible. We numerically verify all of our predictions for system sizes ranging from $N=10^2$ to $N=10^{28}$ and, remarkably, see close agreement over this entire range.

\begin{figure}[t]
\includegraphics[width=\columnwidth]{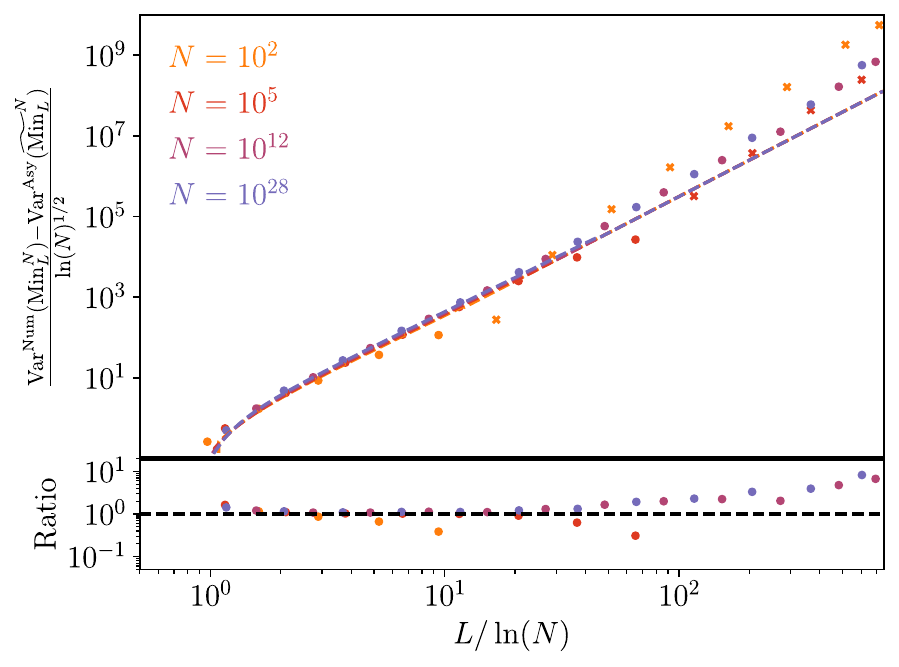}
\caption{We plot $\varnum{\min}-\varasy{\minSSRW}$ for $N=10^2, 10^5, 10^{12}$ and $10^{28}$,  averaged over each horizontal $1/4$ decade. The dots represent positive values of this difference whereas the $\times$ represents the magnitude of negative values of this difference. The dashed line is $\varasy{\env}$ in Eq. \ref{eq:FullEnvVariance} for each $N$ corresponding to its respective color.}
\label{fig:EnvRecoveryAsym}
\end{figure}

Our results point to a potential experimental approach to probe the nature of a random or disordered environment by observing the extreme behavior of many particles diffusing within it. In Figure \ref{fig:EnvRecoveryAsym} we show that it is possible to recover, over multiple decades, the $L^3$ power-law behavior of $\var{\env}$ by numerically measuring $\var{\min}$ and then subtracting the asymptotic theory formula for $\var{\minSSRW}$, the SSRW extreme first passage time variance. This is because the asymptotic behavior of $\var{\minSSRW}$ for $L\gg \ln(N)$ essentially matches that of $\var{\sam}$ in the \rwre{}  model. For the SSRW model (or its Brownian analog), a formula for $\var{\min}$ can be determined asymptotically just by knowing the Einstein diffusion coefficient, which in turn can be estimated experimentally by following the motion of a single particle. Thus, by observing the motion (i.e., extreme first passage time, and Einstein diffusion coefficient) of many particles diffusing in a common environment we are (at least in our numerical simulations) able to recover the power-law that describes the environmental fluctuations. Moreover, with some error, we are also able to estimate the pre-factor of this power-law, which contains further information about the random environment. This pre-factor could be termed the {\it extreme diffusion coefficient} and in future work we plan to develop a more general map between random environments (beyond the special Uniform on $[0,1]$ choice) and extreme diffusion coefficients.

In experimental systems such as $N$ colloids or fluorescent dyes diffusing in quasi-1D channels, or photons (here $N$ relates to the laser intensity) diffusing through a quasi-1D tubes filled with scattering media, it should be possible to precisely observe first passage times to various distances, as well as to estimate the Einstein diffusion coefficient for a single particle. Our numerical conclusions suggest a possible route to observe the hidden environment within which these real diffusions occur. Uncovering an $L^3$ power-law would strongly suggest that the \rwre{} model more accurately captures the behavior of extreme behavior in many particle diffusions. Moreover, the pre-factor to this power-law would constitute a measurement of the {\it extreme diffusion coefficient} and thus serve as a microscope through which to view the hidden environment.

\section{Acknowledgments}
This work was funded under the W.M. Keck Foundation Science and Engineering grant on “Extreme Diffusion".  I.C. also wishes to acknowledge ongoing support from the NSF through DMS:1811143, DMS:1937254, and DMS:2246576, and the Simons Foundation through a Simons Investigator Grant (Award ID 929852). E.I.C. wishes to acknowledge ongoing support from the Simons Foundation for the collaboration Cracking the Glass Problem via Award No. 454939. We wish to thank S. Prolhac for providing data for the numerically computed density of the one-point distribution for the narrow wedge solution to the KPZ equation that was utilized in producing some of the asymptotic curves herein. We also wish to thank P. Le Doussal for helpful comments on a draft of this paper. This work benefited from access to the University of Oregon high performance computing cluster, Talapas.

\bibliographystyle{unsrt}
\bibliography{main}

\end{document}